\documentclass[journal]{IEEEtran}
\usepackage{cite,url}
\usepackage{amsmath,graphicx,epstopdf,amsthm,amssymb,float,color,array,epsfig,subcaption}
\newtheorem{Theorem}{Theorem}

\newtheorem{proposition}{Proposition}
\newtheorem{example}{Special case}
\newtheorem{problem}{Problem}
\newtheorem{definition}{Definition}

\usepackage{algorithm,algorithmic}
\renewcommand{\algorithmicrequire}{\textbf{Input:}}
\renewcommand{\algorithmicensure}{\textbf{Output:}}
\newcommand{\ud}{\,\mathrm{d}}
\newcommand{\avec}{{\bf{a}}}

\newcommand{\evec}{{\bf{e}}}

\newcommand{\yvec}{{\bf{y}}}

\newcommand{\wvec}{{\bf{w}}}
\newcommand{\xvec}{{\bf{x}}}

\newcommand{\vvec}{{\bf{v}}}
\newcommand{\gvec}{{\bf{g}}}

\newcommand{\etavec}{{\bf{\eta}}}

\newcommand{\onevec}{{\bf{1}}}
\newcommand{\zerovec}{{\bf{0}}}

\newcommand{\Lambdamat}{{\bf{\Lambda}}}

\newcommand{\Amat}{{\bf{A}}}
\newcommand{\Bmat}{{\bf{B}}}
\newcommand{\Cmat}{{\bf{C}}}
\newcommand{\Dmat}{{\bf{D}}}
\newcommand{\Emat}{{\bf{E}}}

\newcommand{\Gmat}{{\bf{G}}}

\newcommand{\Mmat}{{\bf{M}}}
\newcommand{\Jmat}{{\bf{J}}}

\newcommand{\Imat}{{\bf{I}}}
\newcommand{\Lmat}{{\bf{L}}}
\newcommand{\Pmat}{{\bf{P}}}

\newcommand{\Qmat}{{\bf{Q}}}

\newcommand{\Smat}{{\bf{S}}}

\newcommand{\Umat}{{\bf{U}}}
\newcommand{\Vmat}{{\bf{V}}}
\newcommand{\Wmat}{{\bf{W}}}

\newcommand{\define}{\stackrel{\triangle}{=}}

\def\bsigma{{\mbox{\boldmath $\Sigma$}}}

\def\bepsilon{{\mbox{\boldmath $\epsilon$}}}

\def\psivec{{\mbox{\boldmath $\psi$}}}

\def\nuvec{{\mbox{\boldmath $\nu$}}}

\def\etavec{{\mbox{\boldmath $\eta$}}}

\def\thetavec{{\mbox{\boldmath $\theta$}}}
\def\thetavecsc{{\mbox{\boldmath \tiny $\theta$}}}

\def\thetavecsmall{{\mbox{\boldmath {\scriptsize $\theta$}}}}

\newcommand{\be}{\begin{equation}}
\newcommand{\ee}{\end{equation}}
\newcommand{\beqna}{\begin{eqnarray}}
\newcommand{\eeqna}{\end{eqnarray}}

\begin{document}
\title{Non-Bayesian Estimation Framework for Signal Recovery on Graphs}
\author{
 Tirza~Routtenberg,~\IEEEmembership{Senior~Member,~IEEE}
\thanks{This research was partially supported by THE ISRAEL SCIENCE FOUNDATION (grant No. 1173/16).  }
\thanks{{\footnotesize{ T. Routtenberg is with the School of Electrical and Computer Engineering Ben-Gurion University of the Negev Beer-Sheva 84105, Israel, e-mail: tirzar@bgu.ac.il.}}}
}
	
	\maketitle
	\nopagebreak
	\begin{abstract}
	Graph signals  arise from physical networks, such as power and communication systems, or as a result of a convenient  representation of
data with complex structure, such as social networks.
We consider the problem of general graph signal recovery  from noisy, corrupted, or incomplete measurements and under  structural parametric constraints, such as smoothness in the graph frequency domain.
In this paper,  we formulate the graph signal recovery as a non-Bayesian estimation problem under 
 a  weighted mean-squared-error (WMSE) criterion, which is based on a quadratic form of the Laplacian matrix of the graph, and its
 trace WMSE is 
the Dirichlet energy of the estimation error w.r.t. the graph. 
 The Laplacian-based WMSE penalizes estimation errors according to their graph spectral content and is a  difference-based cost function which accounts for
the fact that in many cases   signal recovery on graphs can only be achieved up to a constant  addend.
 We  develop  a new Cram$\acute{\text{e}}$r-Rao bound (CRB) 
on the Laplacian-based WMSE and present the associated Lehmann unbiasedness condition w.r.t. the graph.
We discuss 
 the graph CRB and estimation methods for the fundamental problems of
 1) a linear Gaussian  model with relative measurements; and
 2) bandlimited graph signal recovery. We develop sampling allocation policies that optimize sensor locations in a network  for these problems based
on the proposed graph CRB. 
 Numerical simulations on random graphs and on electrical network data are used to validate the performance of the graph CRB and sampling policies.
	\end{abstract}
	\section{Introduction}
	\label{sec:intro}
Graphs are fundamental mathematical structures that are widely used
in various fields for network data analysis to model
 complex
relationships within and between data, signals, and processes. 
Many complex systems in engineering, physics, biology, and sociology constitute networks of interacting units that result in signals that are supported on irregular structures and, thus, can be modeled as signals over the vertices (nodes) of a graph \cite{Newman_2010}, i.e. {\em{graph signals}}.
Thus, graph signals arise in many modern applications, leading to the emergence of the area
 of graph signal processing (GSP) in the last decade
	(see, e.g. \cite{Shuman_Ortega_2013,Sandryhaila_Moura_2013,Chen_Kovavic_2016,8347162}).
	GSP theory  extends concepts and techniques from traditional
digital signal processing (DSP) to data indexed by generic graphs.
 However, most of the research effort in this field has been devoted to the
purely deterministic setting, while methods
that exploit statistical information generally lead to better average performance compared to deterministic methods and are better suited to describe practical scenarios that involve
uncertainty and randomness \cite{segarra2018statistical,7352352}.
In particular, 
the development of performance bounds, such as the well-known  Cram$\acute{\text{e}}$r-Rao bound (CRB), on various estimation problems that are related to the graph structure is a fundamental step towards attaining statistical GSP  tools.

Graph signal recovery aims to recover  graph signals from noisy, corrupted, or incomplete measurements. Applications include  registration of data across a sensor network,  time synchronization across  distributed networks \cite{singer2011three,4177758}, and state estimation in power systems 
\cite{Giannakis_Wollenberg2013,drayer2018detection,GlobalSIP_Drayer_Routtenberg,Grotas_YI_Routtenberg}.
In regular domains, signal recovery is usually performed based on the well-known mean-squared-error (MSE) criterion. However, it is recognized in the literature that the MSE criterion may be limited for characterizing the estimation performance of parameters defined on irregular domains \cite{wang2009mean,Jia_Benson_2019}, such as parameters that are defined on a manifold \cite{8574909,PCRB_J} and in image processing \cite{wang2009mean}. In particular, new cost functions are required for the irregular graph signal recovery \cite{wang2009mean}. This is mainly because an implicit assumption behind the widely-used MSE is that signal fidelity is independent of spatial relationships between the samples of the original signal. In other words, if the original and estimated graph signals are randomly reordered in the same way (i.e. the signal values are rearranged randomly over the graph vertices), then the MSE between them will be unchanged. Apparently, this assumption does not hold for graph signals, which are highly structured. For graph signals, the ordering of the signal samples carries important structural information. As a concrete example of this structural information, anomalies in graph signals are measured w.r.t. their locations and local behavior \cite{Sandryhaila_Moura_2014,drayer2018detection,GlobalSIP_Drayer_Routtenberg}, and, thus, cannot be recognized after reordering the signals.  Similarly, the conventional CRB, which is a lower bound on the MSE, does not display consistency with the geometry of the data \cite{nielsen2013cramer}. 
Moreover, the MSE and the CRB  do not take into account constraints that stem from the graph structure, such as the graph signal smoothness and bandlimitedness w.r.t. the graph, and ignore
the connectivity of the graph and the  degrees of the different nodes, treating the information in connected and separated nodes equally.   Furthermore, in many cases, graph signals  are only a function of relative values, i.e. the differences between vertex values. Such signals arise, for example, in community detection \cite{Newman_2010},  motion consensus \cite{Barooah_Hespanha_2007,Barooah_Hespanha_2008}, time synchronization in networks \cite{singer2011three,4177758}, and power system state estimation (PSSE) \cite{Abur_book,Ariel_Yonina_GlobalSIP}.  In these cases, signal recovery on graphs can only be achieved up to a constant addend, which is a situation which the MSE criterion and its associated CRB  do not address. Therefore, new evaluation measures and performance bounds are required for statistical GSP.
New performance bounds 
can also be useful for
 designing the sensing network topology, which  is a crucial task in  both  data-based and physical networks.

Sampling and recovery of graph signals are fundamental tasks in GSP that have received considerable attention
recently.
In particular, recovery with a regularization using
the Dirichlet energy, i.e. the Laplacian quadratic form,  has been used in various applications, such as image processing \cite{Zheng_Cai2011,Elmoataz_2008}, non-negative matrix factorization
\cite{cai2010graph},  principal component analysis (PCA) \cite{shahid2015robust}, data classification \cite{wang2007label,belkin2004regularization},
and  semisupervised learning on graphs \cite{SSL,Jia_Benson_2019}.
Characterization of graph signals, dimensionality reduction,
  and recovery  using the graph Laplacian as a regularization have been used in various fields \cite{Shuman_Ricaud_Vandergheynst_2012},
  \cite{anis2016efficient}.
 In the case of isotropic Gaussian noise with relative measurements,
the CRB for the synchronization of rotations has been developed in the seminal work in \cite{Boumal_2013,Boumal_Singer_2014},
and it is shown to be   proportional to the pseudo-inverse of the Laplacian matrix.
In addition, the
effect of an incomplete measurement graph on the CRB has been shown to be
related to the Laplacian of the graph \cite{Boumal_2013}.
Further, the  eigenvectors
corresponding to the smallest eigenvalues
of the Laplacian matrix have been shown to determine the CRB in bandlimited graph sampling \cite{anis2016efficient}. 
Thus, the graph Laplacian represents the information on the structure of the underlying graph and
can be used for the development of  performance assessment, analysis, and practical inference
 tools  \cite{rosenberg1997laplacian}.

	In this paper, we study the problem of graph signal recovery in the context of non-Bayesian estimation theory.
First, we introduce   the Laplacian-based weighted MSE (WMSE) criterion as an estimation performance measure for graph signal recovery. 
This measure is used  to quantify  changes w.r.t. the variability that is encoded by the weights of the graph \cite{rudin1992nonlinear} and is a difference-based criterion whose trace is the  Dirichlet energy.
We show that the WMSE can be interpreted as the MSE in the graph frequency domain, in the GSP sense.
We then present the concept of graph-unbiasedness in the sense of the Lehmann-unbiasedness definition \cite{Lehmann}. We develop  a new  Cram$\acute{\text{e}}$r-Rao-type lower bound on the graph-MSE (Laplacian-based WMSE) of any graph-unbiased estimator. 
The proposed graph CRB is examined  for a linear Gaussian model  with relative measurements and for bandlimited graph signal recovery.
We show that the new bound  provides analysis and design tools, where we
optimize the sensor locations by using the graph CRB.
In simulations, we demonstrate the use of the graph CRB  and associated estimators for recovery of graph signals in random graphs and  for the problem of PSSE in electrical networks. In addition,  we show that the proposed sampling policies lead to better estimation performance in terms of Dirichlet energy.


	The rest of the paper is organized as follows:
	Section \ref{probFor} presents the mathematical model for the  graph signal recovery.
	In Section \ref{GMSE_sec} 	we present the proposed performance measure 
	and  discuss its properties. 
	In Section \ref{sCRB} the new graph CRB is derived.
	In Section \ref{ex1} and Section \ref{Bandlim_subsection}, we develop the graph CRB for a
	linear Gaussian model with relative measurements
	and  for	bandlimited graph signal recovery, respectively, and discuss the design of sample allocation policies based on the new bound. 
	The performance of the proposed graph CRB  is evaluated in simulations in Section \ref{simulation_sec}.
	Finally, our conclusions can be found in Section \ref{conclusion}.

\section{Model and problem formulation}
\label{probFor}
In this section, we present the considered model and formulate the graph signal recovery task as a non-Bayesain estimation problem.
Subsection \ref{notation_sec} includes the notations used in this paper.
Subsection \ref{GSP_sec} introduces the background and relevant concepts of
GSP. In Subsection \ref{cost_sec} and in Subsection \ref{const_subsec}
we present the estimation problem and the influence of linear parametric constraints, respectively.

\subsection{Notations}
\label{notation_sec}
	In the rest of this paper, we denote vectors by boldface lowercase letters and matrices by boldface uppercase letters. 
	The operators  $(\cdot)^T$, $(\cdot)^{-1}$, $(\cdot)^{\dagger}$, and ${\text{Tr}}(\cdot)$ denote the transpose, inverse,  Moore-Penrose pseudo-inverse, and trace operators, respectively.
For a matrix $\Amat\in{\mathbb{R}}^{M\times K}$ with a full column rank,
		$\Pmat_\Amat^\bot=\Imat_M-\Amat(\Amat^T\Amat)^{-1}\Amat^T$,
		where $\Imat_M$ is the  identity matrix of order $M$.
		The matrix  ${\text{diag}}(\avec)$ denotes the diagonal matrix with vector $\avec$ on the diagonal.
	The $m$th element of the vector $\avec$ and
	the $(m,q)$th element of the matrix $\Amat$
		are denoted by $a_m$ and $\Amat_{m,q}$,  respectively. 
		 Similarly, $\Amat_{\mathcal{S}_1,\mathcal{S}_2}$
		 is used to denote  the submatrix of $\Amat$ 
	 whose rows are indicated by  the  set $\mathcal{S}_1$
	 and columns are indicated by  the  set $\mathcal{S}_2$.  For simplicity, we denote $\Amat_{\mathcal{S},\mathcal{S}}$ by $\Amat_{\mathcal{S}}$.
	For two symmetric  matrices  of the same size
	$\Amat$ and $\Bmat$, $\Amat\succeq\Bmat$ means that $\Amat-\Bmat$ is a positive semidefinite matrix.
	The gradient of a vector function $\gvec(\thetavec)$, $\nabla_{\thetavecsmall}\gvec(\thetavec)$, is a matrix in $\mathbb{R}^{K\times M}$, with the $(k,m)$th element equal to $\frac{\partial g_k}{\partial \theta_m}$, where $\gvec=\left[g_1,\ldots,g_K\right]^T$ and $\thetavec=\left[\theta_1,\ldots,\theta_M\right]^T$. For any index set, ${\mathcal{S}} \subset \{1,\dots,M\}$,
	$\thetavec_{{\mathcal{S}}}$ is a subvector of $\thetavec$ containing the elements indexed by ${\mathcal{S}}$, where  $|{\mathcal{S}}|$ and ${\mathcal{S}}^c \define \{1,\dots,M\}\backslash{\mathcal{S}}$ denote the set's cardinality and the complement set, respectively. 
The vectors $\onevec $ and $\zerovec$   are column vectors of ones and zeros, respectively, $\evec_m$  is the $m$th column of the identity matrix, all with appropriate dimensions. The notation 
$\mathbf{1}_\mathcal{A}$ denotes the indicator  of an event $\mathcal{A}$. The number of non-zero entries in $\thetavec$ is denoted by $||\thetavec||_0$.  
Finally, 
the notation  ${\rm{E}}_{\thetavecsmall} [\cdot]$  represents the  expected value  parameterized by a deterministic parameter $\thetavec$.

\subsection{Background: Graph signal processing (GSP)}
\label{GSP_sec}
In this section, we briefly review relevant concepts related to GSP  \cite{Shuman_Ortega_2013,Sandryhaila_Moura_2013} that will be used in this paper.
Consider an  undirected weighted graph,  ${\mathcal{G}}({\mathcal{M}},\xi,\Wmat)$,
where ${\mathcal{M}}=\left\{1,\ldots,M\right\}$ denotes the set of $M$ nodes or vertices and
$\xi$ denotes the set of edges with cardinality $|\xi|$.
We only consider simple graphs, with no self-loops and multi-edges.
The symmetric matrix  $\Wmat$ is the weighted adjacency
matrix
with entry $\Wmat_{m,k}$ denoting the weight
of the edge $(m,k)\in \xi$, reflecting the 
strength of the connection  between the
nodes $m$ and $k$. This weight may be a physical measure or conceptual, such as a similarity measure. We assume for simplicity
that the edge weights in $\Wmat$ are non-negative ($\Wmat_{m,k}\geq 0$).
When no edge exists between $m$ and $k$, the weight
is set to 0, i.e. $\Wmat_{m,k}= 0$.
\begin{definition}
\label{neighborhood}
 Given ${\mathcal{G}}({\mathcal{M}},\xi,\Wmat)$,
the neighborhood of a node $m\in {\mathcal{M}}$ is defined as
${\mathcal{N}}_m = \{k \in {\mathcal{M}} : (m,k) \in \xi\}$.
\end{definition}


The Laplacian matrix, which  contains the information
on the graph structure, is defined by   $\Lmat \define \Dmat-\Wmat$, where $\Dmat$ is a diagonal matrix with $\Dmat_{m,m}=\sum_{k=1}^M\Wmat_{m,k}$.
The Laplacian matrix, $\Lmat$, is a real, symmetric, and positive semidefinite matrix,  which satisfies
the null-space property, $\Lmat\onevec_M=\zerovec$,
and has nonpositive off-diagonal elements. 
Thus, its associated singular value decomposition (SVD) is given by 
 \begin{equation}
\label{SVD_new_eq}
\Lmat = \Vmat\Lambdamat \Vmat^T,   
 \end{equation}
where the columns of $\Vmat$ are the eigenvectors of $\Lmat$, $\Vmat^T=\Vmat^{-1}$, and $\Lambdamat \in \mathbb{R}^{M \times M}$ is a diagonal matrix consisting of the distinct  eigenvalues of 
$\Lmat$, $0= \lambda_1 < \lambda_2 < \ldots < \lambda_M $. 
Throughout this paper we will focus
on the case where the observed graph is connected and, thus, $\lambda_m> 0$, $m=2,\ldots,M$. If the
graph is not connected, 
the proposed approach can be applied to each connected component separately.
The eigenvalues
$\lambda_1,\ldots,\lambda_M$ can be interpreted as graph frequencies, and eigenvectors, i.e. the columns of  the matrix 
$\Vmat$,  can be interpreted as corresponding graph frequency components.
Together they define the graph spectrum for graph  ${\mathcal{G}}({\mathcal{M}},\xi,\Wmat)$.

In this framework, a {\em{graph signal}} is defined
as a function $\thetavec:{\mathcal{M}}\rightarrow {\mathbb{R}}^M$,  assigning a scalar value to each vertex, where entry $\theta_m$ denotes the signal value at node $m\in {\mathcal{M}}$.
The graph Fourier transform (GFT)  of a graph signal $\thetavec$ w.r.t. the graph ${\mathcal{G}}({\mathcal{M}},\xi,\Wmat)$ is defined as the projection onto the
orthogonal set of the eigenvectors of $\Lmat$  \cite{Shuman_Ortega_2013,Sandryhaila_Moura_2013}:
 \begin{equation}
\label{GFT}
\tilde{\thetavec}\triangleq \Vmat^T\thetavec.  
 \end{equation}
 Similarly,  the inverse GFT is obtained by left multiplication of a vector by $\Vmat$, i.e. by $\Vmat\tilde{\thetavec}$.
The GFT
plays a central role in GSP since it is a
natural extension of filtering operations and the notion of  the spectrum of the graph signals \cite{Sandryhaila_Moura_2013,Shuman_Ortega_2013}.

\subsection{Estimation problem}
\label{cost_sec}
We consider the problem of estimating the graph signal, $\thetavec\in {\mathbb{R}}^M$, which is considered in this paper to be a deterministic parameter vector.
The estimation is based on a random  observation vector, $\xvec \in \Omega_\xvec$, where $\Omega_\xvec$ is a general observation space.
We assume that $\xvec$ is distributed according to a known probability distribution function (pdf), $f(\xvec;\thetavec)$,
which is parametrized by the graph signal, $\thetavec$. For example,
for the problem of mean estimation
   in
Gaussian graphical models (GGM)
\cite{deshpande2004model,Wiesel_Hero_2012},
$f(\xvec;\thetavec)$ is a Gaussian distribution with mean $\thetavec$ and  covariance matrix, $\bsigma$, where $\bsigma^{-1}$ preserves the connectivity pattern of the graph.
Our goal is to integrate the side information in the form of graph structure, encoded by the Laplacian matrix, in the estimation approach.

Let  $\hat{\thetavec} :\Omega_\xvec\rightarrow{\mathbb{R}}^M$ be an estimator of $\thetavec$, based on a random observation vector, $\xvec\in\Omega_\xvec$. We consider  estimators in the  Hilbert space of 
bounded energy  on ${\mathcal{G}}({\mathcal{M}},\xi,\Wmat)$
 \cite{bezuglyi2018graph},
i.e. we assume that $
 {\rm{E}}_{\thetavecsmall}[\hat{\thetavec}^T\Lmat \hat{\thetavec}]
<\infty$.
Our goal in this paper is to investigate estimation methods and bounds that are based on defining a structural estimation  performance measure that reflects the graph topology and graph signal properties.

\subsection{Linear constraints in graph recovery}
\label{const_subsec}
In various GSP problems there is side information on the graph signals in the form of
 linear parametric constraints.
These linear constraints describe properties of the graph signals, such as bandlimitedness (see the example in Subsection \ref{Bandlim_subsection}), or properties of the sensing approach, such as the existence of reference nodes (anchors) \cite{Barooah_Hespanha_2007,Barooah_Hespanha_2008}, 
 can serve to obtain a  well-posed estimation problem \cite{Boumal_2013}.
Formally,
in these cases it is known {\em{a-priori}} that $\thetavec$ satisfies the following linear constraint:
\be\label{linear}
\Gmat\thetavec+\avec=\zerovec,~\Gmat\in\mathbb{R}^{K\times M},
\ee
where $\Gmat\in{\mathbb{R}}^{K\times M}$ and $\avec\in\mathbb{R}^{K}$ are known,
and $0\leq K\leq M$. We assume that the matrix $\Gmat$  
has full row rank,  i.e. the constraints are not redundant.
The constrained set is: 
\be \label{set_def}
\Omega_\thetavecsmall\define\{\thetavec\in{\mathbb{R}}^M|\Gmat\thetavec+\avec=\zerovec\}.
\ee
Thus, we are interested in the problem of recovering graph signals that belong to $\Omega_\thetavecsmall$.
We define  the orthonormal null space matrix,
$\Umat\in{\mathbb{R}}^{M\times(M-K)}$, such that \cite{Hero_constraint}
\be \label{one}
\Gmat\Umat=\zerovec {\text{ and }}
\Umat^T\Umat=\Imat_{M-K}.
\ee
The case $K=0$ represents an unconstrained estimation problem, in which we use the convention that $\Umat=\Imat_M$ \cite{Stoica_Ng,Eyal_constraint,sparse_con}.
The matrix $\Umat$ can be found 
based on the eigenvector matrix of 
the orthogonal projection matrix $\Pmat_\Gmat^\perp \define \Imat_M- \Gmat^T(\Gmat\Gmat^T)^{-1}\Gmat$.
\subsection{Motivating example: State estimation in power systems}
\label{sub_sec_motivation}
\begin{figure}[htb]
         \begin{minipage}{.42\linewidth}
     \centering
\subcaptionbox{\label{fig01}}[\linewidth]
{\includegraphics[width=4cm]{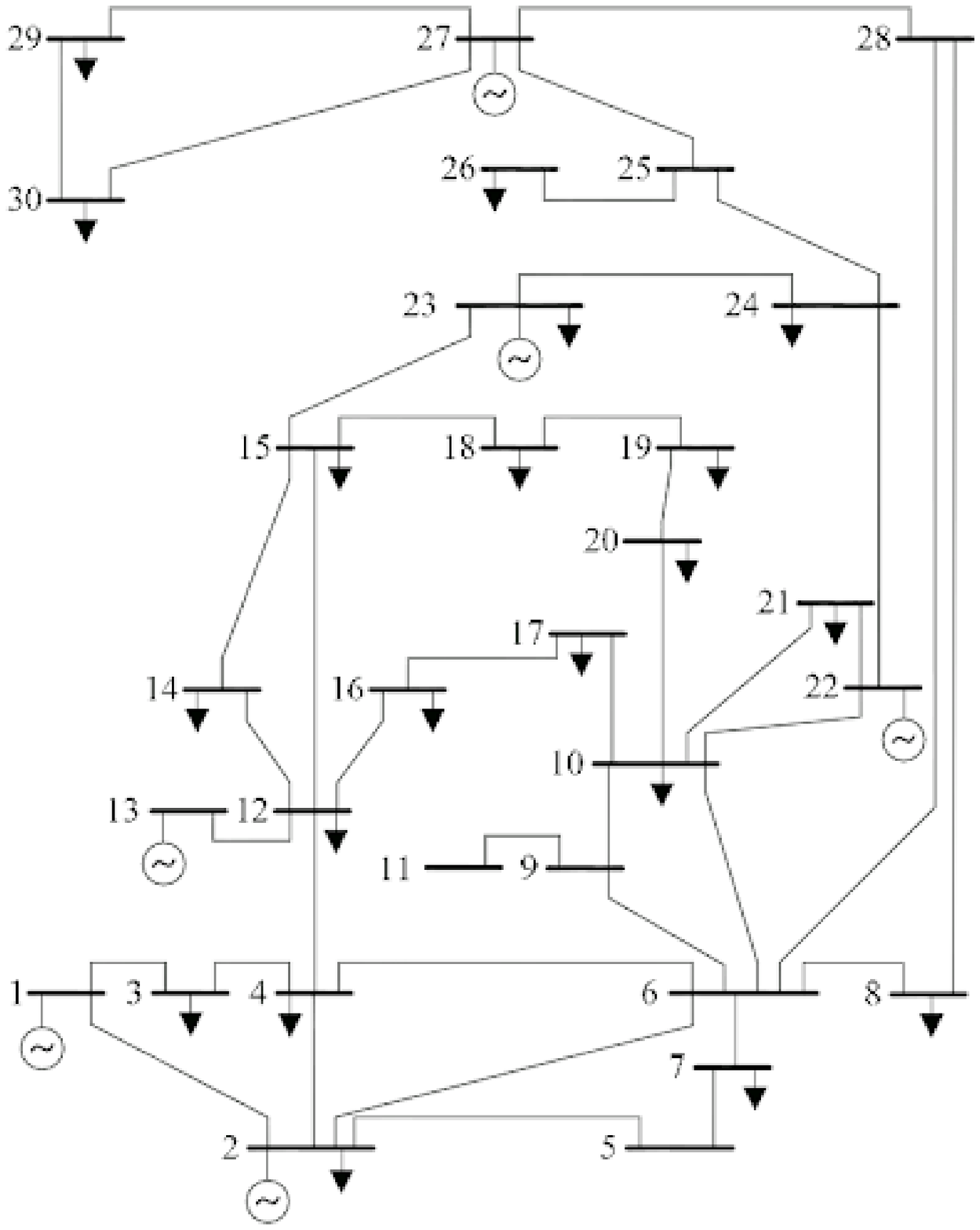}}
\end{minipage}
\hspace{-0.1cm}
    \begin{minipage}{.49\linewidth}
\centering
\subcaptionbox{\label{fig02}}[\linewidth]
{ \includegraphics[width=4.5cm]{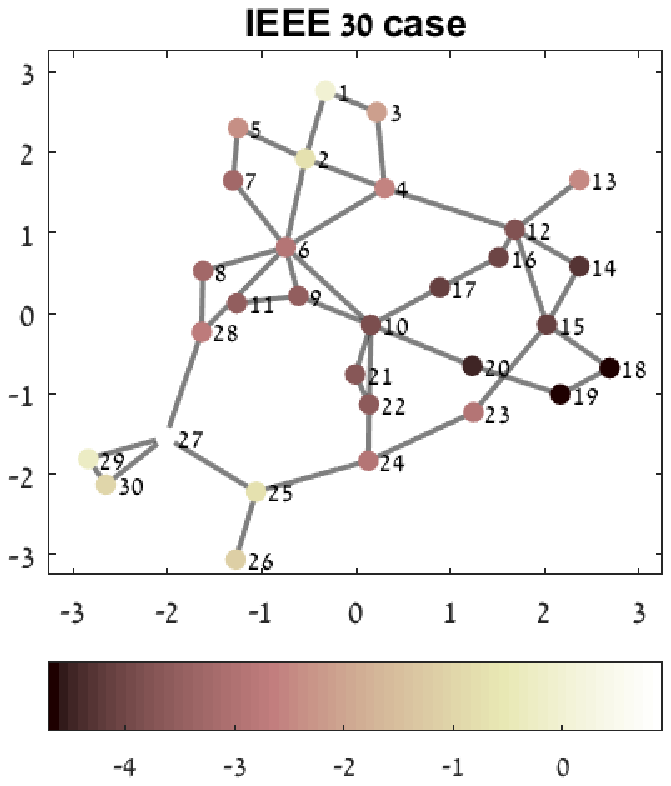}}
  \end{minipage}
\vspace{-0.1cm}
     \caption{Motivating example: IEEE 30-bus system \cite{iEEEdata} (a)  and  its representation as a graph (b). The node
color represents the state signal value.}
 \label{Fig0}
 \vspace{-0.25cm}
 \end{figure}

This paper deals with a general parameter estimation over networks. In order to demonstrate how our proposal can be applied  in practice, 
 we give a physically-motivated example of state estimation in power systems.
A power system can be represented as an undirected  weighted graph,
${\mathcal{G}}({\mathcal{M}},\xi,\Wmat)$, where the set of vertices, $\mathcal{M}$, is the set of buses (generators or loads) and the edge set, $\xi$, is the set of transmission lines between these buses. The weight matrix, $\Wmat$, is determined by the branch susceptances \cite{Grotas_YI_Routtenberg,Ariel_Yonina_GlobalSIP}.
The goal of PSSE, which is at the core  of energy management systems for various monitoring and analysis purposes,   is to estimate $\thetavec$ based on system measurements, $\xvec$,
that usually include power measurements
\cite{Abur_book}.
Since $\thetavec$ is measured over the buses of the electrical network,
PSSE can be interpreted  as a graph signal recovery problem.
The pdf,  $f(\xvec;\thetavec)$, describes the physical relation between the power and the voltages, as well as the statistical behavior of noises due to, e.g. errors and varying temperatures.
In
Fig. \ref{Fig0}.a, we show the one-line diagram of the
IEEE 30-bus system, which is a well-known test grid for power
system applications \cite{iEEEdata}. In Fig. \ref{Fig0}.b, we show the graphical representation
of this grid. The node
color represents the state signal value. It can be seen that neighboring buses have a similar value (similar color) and, thus, the state signal is smooth.

The characteristics of the state estimation problem in power systems are as follows. First, the power flow equations are an
up-to-a-constant function w.r.t. the state vector, $\thetavec$ \cite{Ariel_Yonina_GlobalSIP}, i.e. relative measurements, as described in Section \ref{ex1}.
Thus, the MSE cannot be used directly as a performance measure without the use of    a reference bus  (node).
Additionally, 
it has recently been shown in
 \cite{GlobalSIP_Drayer_Routtenberg,drayer2018detection} that the state vector in power systems   is a smooth graph signal w.r.t. the associated  Laplacian matrix that is determined by the susceptances values. 
Thus, the GSP framework is well suited for PSSE, as shown, for example, in \cite{drayer2018detection,GlobalSIP_Drayer_Routtenberg} for addressing the problem of detection of   false data injection attacks in power systems based on the GFT of the states.

	\section{Laplacian-based WMSE estimation}
	\label{GMSE_sec}
Sampling and recovery of graph signals are fundamental
tasks in GSP that have received considerable attention.
In regular domains, signal recovery is usually performed based on the  MSE criterion.
 However,  the MSE  may be limited for characterizing the performance of  graph signal recovery since it ignores the structural relationship in the neighborhood of the graph signals \cite{wang2009mean,Jia_Benson_2019}.
In this section, we
suggest the use of the  WMSE for graph signal recovery.
The proposed WMSE measures the variation of the estimation error w.r.t. the underlying graphs.
The rationale behind  the WMSE
as a difference-based criterion that fits various applications, its Dirichlet  energy  interpretation, and its graph frequency interpretation are discussed in Subsections \ref{Dif_Subs}, \ref{TVsubsec}, and  \ref{GFT_interpretation}, respectively.

In this paper, we
suggest the use of the following  
matrix cost function for graph signal recovery:
\beqna
\label{Cmat_def}
\Cmat(\hat{\thetavec},\thetavec)\define
\Lambdamat^{\frac{1}{2}}\Vmat^T (\hat{\thetavec}-\thetavec)(\hat{\thetavec}-\thetavec)^T\Vmat\Lambdamat^{\frac{1}{2}},
\eeqna
where $\Vmat$ and $\Lambdamat$ are defined in \eqref{SVD_new_eq}.
The corresponding risk, which  is  the expected cost from \eqref{Cmat_def}, is the following WMSE:
	\beqna
	\label{MSSE}
		{\rm{E}}_{\thetavecsmall}[\Cmat(\hat{\thetavec},\thetavec)]
		=\Lambdamat^{\frac{1}{2}} \Vmat^T{\text{MSE}}(\hat{\thetavec},\thetavec)\Vmat
		\Lambdamat^{\frac{1}{2}},
	\eeqna
where the MSE matrix is defined as:
\beqna
\label{MSE}
  {\text{MSE}}(\hat{\thetavec},\thetavec)\define 
{\rm{E}}_{\thetavecsmall}[(\hat{\thetavec}-\thetavec)(\hat{\thetavec}-\thetavec)^T].
\eeqna
The proposed  risk in \eqref{MSSE}
 can be interpreted as a WMSE criterion \cite{ELDAR_WEIGHTED_MSE,Eyal_constraint} with a positive semidefinite weighting matrix, $\Lambdamat^{\frac{1}{2}} \Vmat^T$.
 Different weights are
assigned to the individual errors, where these weights are based on the graph topology.
The Laplacian matrix, $\Lmat$, which is used to determine these weights through
$\Lambdamat$ and $\Vmat^T$,
can be based on a physical network or on statistical dependency, such as a probabilistic graphical model that is  obtained from  historic, offline data.

The rationale behind this cost function, as well as some of its properties, are as follows:
\subsubsection{Difference-based criterion}
\label{Dif_Subs} 
It is proved in Appendix \ref{appA} that
the $(m,n)$th element of the matrix cost, $\Cmat(\hat{\thetavec},\thetavec)$, is
\beqna
\label{Cmn3}
\Cmat_{m,n}(\hat{\thetavec},\thetavec)\hspace{6cm}
\nonumber\\=\Lambdamat^{\frac{1}{2}}_{m,m}  \Lambdamat^{\frac{1}{2}}_{n,n}\sum_{k=1}^M\sum_{l=1}^M
\Vmat_{k,m} \Vmat_{l,n}
(\epsilon_k-\epsilon_m)(\epsilon_l-\epsilon_n),
\eeqna
where the estimation error at node $m$ is defined by
\be
\label{error_eps}
\epsilon_m\define \hat{\theta}_m-\theta_m,~m=1,\ldots,M.
\ee
The elements of the cost matrix in \eqref{Cmn3} imply that the proposed matrix cost function
only takes into account the relative estimation errors of 
the differences between the error signals at different nodes.
 Thus, any translation of all errors by a vector $c\onevec$,
 would not change the matrix cost.
This property reflects the fact that in various applications,
such as control, power systems, and image processing 
 \cite{Barooah_Hespanha_2007,Barooah_Hespanha_2008,Jia_Benson_2019,singer2011three,4177758,Abur_book},
graph signals are only a function of the differences between vertex values,  
and
estimation can only be achieved up to a constant  addend \cite{Ariel_Yonina_GlobalSIP,Barooah_Hespanha_2007,Barooah_Hespanha_2008}.
It should be noted that the derivation of \eqref{Cmn3} in Appendix \ref{appA} is based on the fact that  $(\lambda_1,\vvec_1)=(0,\frac{1}{\sqrt{M}}\onevec)$ is  the 
smallest eigenvalue-eigenvector pair of $\Lmat$. Thus, \eqref{Cmn3} does not hold for other forms of the Laplacians.

 In addition, since for the Laplacian matrix $\lambda_1=0$,
then $\Lambdamat_{1,1}=\lambda_1=0$. By substituting this value in \eqref{Cmn3},
we obtain
\be
\label{12}
\Cmat_{m,n}(\hat{\thetavec},\thetavec)=0,
~{\text{if }}m=1
{\text{ and{\slash}or }}n=1.
\ee
Therefore, 
minimization of the expected matrix cost, ${\rm{E}}_\thetavecsmall[\Cmat(\hat{\thetavec},\thetavec)]$, in the sense of positive semidefinite matrices, is equivalent to minimization of the submatrix ${\rm{E}}_\thetavecsmall[\Cmat_{\mathcal{M}\setminus 1}(\hat{\thetavec},\thetavec)]$.
\subsubsection{Dirichlet energy interpretation}
\label{TVsubsec} 
By substituting \eqref{SVD_new_eq} in \eqref{Cmat_def}  and using the trace operator properties, it can be verified that
the trace of $\Cmat(\hat{\thetavec},\thetavec)$ is given by
\beqna
\label{aSSE}
{\text{Tr}}(\Cmat(\hat{\thetavec},\thetavec))
=(\hat{\thetavec}-\thetavec)^T\Vmat\Lambdamat \Vmat^T (\hat{\thetavec}-\thetavec)
\nonumber\\
=(\hat{\thetavec}-\thetavec)^T\Lmat (\hat{\thetavec}-\thetavec).\hspace{0.8cm}
\eeqna
By substituting $\Lmat=\Dmat-\Wmat$ in \eqref{aSSE} it can be shown
that 
\beqna
\label{aSSE2}
{\text{Tr}}(\Cmat(\hat{\thetavec},\thetavec))
=\frac{1}{2}\sum_{m=1}^M\sum_{k \in{\mathcal{N}}_m } \Wmat_{m,k}(\epsilon_k-\epsilon_m)^2,
\eeqna
where ${\mathcal{N}}_m$ is the set of connected neighbors of node $m$, as defined in Definition \ref{neighborhood}.

The trace of the cost in \eqref{aSSE} and \eqref{aSSE2}
is the
Dirichlet energy of  the estimation error signal,
$\bepsilon=\hat{\thetavec}-\thetavec$,  w.r.t. the Laplacian matrix $\Lmat$ \cite{belkin2002laplacian,Shuman_Ricaud_Vandergheynst_2012}.
This smoothness measure, which is well known in spectral graph theory, quantifies how much the signal changes w.r.t. the variability encoded by the graph weights.  Intuitively, since the weights are nonnegative, the graph Dirichlet energy in \eqref{aSSE} shows that an estimator, $\hat{\thetavec}$, is considered to be a ``good estimator" if the error is a smooth signal and the error values are close to their neighbors’ values,
 as described on the right hand side (r.h.s.) of \eqref{aSSE2}.
Additionally,  it can be seen from \eqref{aSSE2} that the proposed measure  penalizes the squared  differences of estimation errors  proportionally to the weights of connections
between them,  $\{\Wmat_{m,k}\}_{(m,k)\in {\xi}}$.
This is due to the fact that the errors in highly-connected vertices   with a large degree have more 
influence on subsequent processing than  those of  separated vertices and, thus, should be penalized more.  


\subsubsection{Graph frequency interpretation}
\label{GFT_interpretation}
By substituting  the GFT operator from \eqref{GFT} in \eqref{Cmat_def},
we obtain the equivalent graph frequency cost function:
\beqna
\label{Cmat_def_frec}
\Cmat(\hat{\thetavec},\thetavec)= \Lambdamat^{\frac{1}{2}} 
(\hat{\tilde{\thetavec}}-\tilde{\thetavec})(\hat{\tilde{\thetavec}}-\tilde{\thetavec})^T\Lambdamat^{\frac{1}{2}},
\eeqna
where $\hat{\tilde{\thetavec}}\define \Vmat^T \hat{\thetavec}$ can be interpreted as an estimator of  $\tilde{\thetavec}$. 
That is, the proposed matrix cost function penalizes the {\em{marginal}} estimation errors in the frequency domain, weighted by the associated eigenvalue.
In particular, since for the Laplacian matrix $\lambda_1=0$, 
then $\Lambdamat_{1,1}=0$ and $\hat{\tilde{\theta}}_1-\tilde{\theta}_1$, does not affect the cost function in \eqref{Cmat_def_frec}. 
By applying the trace operator on \eqref{Cmat_def_frec}, one obtains
\beqna
	\label{aSSE3}
		{\text{Tr}}(\Cmat(\hat{\thetavec},\thetavec))
		=\sum_{m=2}^M \lambda_m (\hat{\tilde{\theta}}_m-\tilde{\theta}_m)^2.
	\eeqna

\subsubsection{Alternative cost function}
\label{alternative_sub_sec}
It should be noted that
the  matrix cost function $\Cmat(\hat{\thetavec},\thetavec)$ from
\eqref{Cmat_def}
can be  straightforwardly modified according to the specific parameter estimation problem. For example, 
 problem dimensionality can  be reduced using only part of the graph frequencies,
 instead of the full Laplacian matrix,
 for the reconstruction of graph bandlimited signals. 
 That is, we can use the cost matrix 
 \be
 \label{cost_bandli}
 \Lambdamat_{\mathcal{S},\mathcal{S}}^{\frac{1}{2}}\Vmat_{\mathcal{M},\mathcal{S}}^T (\hat{\thetavec}-\thetavec)(\hat{\thetavec}-\thetavec)^T
 \Vmat_{\mathcal{M},\mathcal{S}}\Lambdamat_{\mathcal{S},\mathcal{S}}^{\frac{1}{2}},
 \ee
 where $\mathcal{S}\subseteq {\mathcal{M}}$ is a subset of the indices of the graph frequencies. 
In addition, the cost function in \eqref{Cmat_def} can be extended to different  smoothness measures, e.g.  by using 
\be
\Lambdamat^{\frac{p}{2}}\Vmat^T (\hat{\thetavec}-\thetavec)(\hat{\thetavec}-\thetavec)^T\Vmat\Lambdamat^{\frac{p}{2}},~p\geq 1.
\ee
Another class of  matrix costs can be obtained by using the SVD of  other forms of Laplacian, such as the random-walk Laplacian, and any
general graph-shift operator (GSO), $\Smat$, which is
an $M\times M$ matrix whose  $(m,k)$th entry is zero for any nonadjacent vertices,  $m$ and $k$.
For example, in \cite{Jia_Benson_2019}, the sum-of-squares errors between all connected vertices, i.e. a loss function that is based on weighting by the adjacency matrix of the network, has been proposed.
Another alternative cost function can be defined based on  different choices of GFT bases with variation operators, as described in Table 1 in \cite{anis2016efficient}.
In general, taking practical considerations into account, a cost function should capture the relevant errors meaningfully and, at the same time, be easily computed. 
An example for an alternative cost function is described for  bandlimited graph signal recovery in Subsection \ref{alternative_bandli_sub}.

\section{Graph CRB}
\label{sCRB}
The CRB is a commonly-used  lower bound on the MSE matrix from \eqref{MSE} of any unbiased estimator.
	In this section, a  Cram$\acute{\text{e}}$r-Rao-type lower bound for  graph-signal recovery, which is useful for performance analysis and system design in networks,
	is derived in Section \ref{sCRBsec}.
	The proposed  graph CRB is a lower bound on the WMSE from \eqref{MSSE} of any  unbiased estimator, where the unbiasedness w.r.t. the graph is defined in Section \ref{sUnbias} by using the concept of Lehmann unbiasedness.
	\subsection{Graph unbiasedness}
	\label{sUnbias}
	Lehmann \cite{Lehmann} proposed a generalization of the unbiasedness concept based on the chosen cost function for each scenario, which can be used for various cost functions (see, e.g.
	\cite{Routtenberg_Tong_est_after_2016,PCRB_J,PCRB_J,Eyal_constraint}).
The  Lehmann unbiasedness definition for a matrix cost functions is defined  as follows (p. 13 in \cite{RoutPhd}):
	\begin{definition}
\label{unbiased_definition}
		The estimator, $\hat{\thetavec}$, is said to be a uniformly unbiased estimator of $\thetavec$ in the Lehmann sense w.r.t. a general positive semidefinite matrix cost function, $\Cmat(\hat{\thetavec},\thetavec)$, if
		\be
			\label{Lehmann_vector}
			{\rm{E}}_{\thetavecsmall}[\Cmat(\hat{\thetavec},\etavec)] \succeq {\rm{E}}_{\thetavecsmall}[\Cmat(\hat{\thetavec},\thetavec)],~\forall \etavec,\thetavec\in \Omega_\thetavecsmall,
		\ee
		where $\Omega_\thetavecsmall$ is the parameter space.  
	\end{definition}
		The following proposition defines a sufficient condition for
 the  graph unbiasedness  of estimators in the Lehmann sense w.r.t. the  Laplacian-based WMSE and under the linear parametric constraints.
	\begin{proposition}
		\label{propLehm}
		If an estimator, $\hat{\thetavec}$,   
 satisfies
		\beqna
	\label{unbiased_cond}
	\Umat^T \Lmat
	{\rm{E}}_{\thetavecsmall}[ \hat{\thetavec}-\thetavec]=\zerovec, ~\forall \thetavec\in \Omega_\thetavecsmall,
	\eeqna
	where $\Omega_\thetavecsmall$
	is defined in \eqref{set_def}, 
	then, $\hat{\thetavec}$
		is an unbiased estimator of the graph signal, $\thetavec$, 	in the Lehmann sense, as defined in Definition \ref{unbiased_definition},
		w.r.t. the  weighted squared-error cost function from \eqref{Cmat_def} and under the constrained set in \eqref{linear}.
	\end{proposition}
	\begin{IEEEproof} 
	The proof is given in Appendix \ref{unbiasedApp}.
	\end{IEEEproof} 
 It can be seen that if an estimator has a zero mean bias,  i.e. 
${\rm{E}}_{\thetavecsmall}[\hat{\thetavec}]-\thetavec=\zerovec$, $\forall\thetavec\in\Omega_\thetavecsmall$, then it satisfies \eqref{unbiased_cond}, but not vice versa. Thus, the uniform graph unbiasedness condition is a weaker condition than requiring
the   mean-unbiasedness property.
For example, since
$ \Lmat\onevec=\zerovec$,
  the condition in \eqref{unbiased_cond}
is oblivious to the
estimation error of a constant bias over the graph, $c \onevec$,
for any constant $c\in{\mathbb{R}}$, in contrast with mean unbiasedness.
	It can be seen that the graph unbiasedness in (\ref{unbiased_cond}) is  a function of the specific graph topology.  In addition,
 the unbiasedness definition can be reformulated by using any 
matrix   that spans the null space of
$\Gmat$, ${\mathcal{N}}(\Gmat)$, instead of $\Umat$.

\begin{example}
\label{special_case1}
For the  case where no constraint is imposed, i.e. $K=0$,
	 we have $\Umat=\Imat_M$ and the condition in \eqref{unbiased_cond} is reduced to
		\beqna
	\label{unbiased_cond_w_out}
	\Vmat\Lambdamat  
	{\rm{E}}_{\thetavecsmall}
	[ \hat{\tilde{\thetavec}}-\tilde{\thetavec}]=\zerovec, ~\forall \tilde{\thetavec}\in {\mathbb{R}}^M,
	\eeqna
	where we used the GFT operator from \eqref{GFT}.
	 The condition in \eqref{unbiased_cond_w_out} is equivalent to the requirement that  for each nonzero eigenvalue, the bias in the frequency domain should be zero.
	\end{example}
Another special case of the graph-unbiasedness for a bandlimited graph signal is discussed in Subsection \ref{graph_crb_bandlimited}.

In non-Bayesian estimation theory, two types of unbiasedness are usually considered: uniform unbiasedness at any point in the parameter space, and local unbiasedness, in which the estimator is assumed to be unbiased only in the vicinity of the parameter $\thetavec_0$.
By using the  feasible directions of the constraint set, similar to the derivations  in \cite{sparse_con,Eyal_constraint},
it can be shown that the  local graph-unbiasedness conditions  are as follows:
\begin{definition} \label{LocCunbias_prop}
Necessary conditions for an estimator $\hat{\thetavec}:\Omega_\xvec\rightarrow{\mathbb{R}}^M$
to be a locally Lehmann-unbiased estimator in the vicinity of $\thetavec_0\in\Omega_\thetavecsmall$  w.r.t. the WMSE   are
\be \label{point_wise_unb}
\Umat^T \Lmat
\left.	{\rm{E}}_{\thetavecsmall}[ \hat{\thetavec}-\thetavec]\right|_{\thetavecsmall=\thetavecsmall_0}=\zerovec
\ee
and
\be\label{sec_localU_final}
\Umat^T	\Lmat \left.\nabla_{\thetavecsmall}{\rm{E}}_{\thetavecsmall}
	[\hat{\thetavec}-\thetavec]\right|_{\thetavecsmall=\thetavecsmall_0} \Umat=\zerovec.
\ee
\end{definition}

	\subsection{Graph CRB}
	\label{sCRBsec}
In the following, a novel graph CRB for the estimation of graph parameters is derived. 
Various low-complexity, distributive algorithms exist for graph signal recovery. The new bound is a useful tool for assessing their performance.
The proposed graph CRB is a lower bound on the WMSE of local graph-unbiased estimators
	in the vicinity of $\thetavec$.
	\begin{Theorem} \label{T3}
Let $\hat{\thetavec}$ be a locally graph-unbiased estimator of $\thetavec$, 
 as defined in Definition \ref{LocCunbias_prop}, and assume the following regularity conditions: 
\renewcommand{\theenumi}{C.\arabic{enumi}} 
\begin{enumerate}
\setcounter{enumi}{0}
\item \label{cond2} 
The operations of integration w.r.t. $\xvec$ and differentiation w.r.t. $\thetavec$ can be interchanged, as follows:
		\beqna
			\nabla_{\thetavecsmall}\int_{\Omega_\xvec}\gvec(\xvec,\thetavec)\ud\xvec
			=\int_{\Omega_\xvec}\nabla_{\thetavecsmall}\gvec(\xvec,\thetavec)\ud\xvec,
		\eeqna
	 for any measurable function $\gvec(\xvec,\thetavec)$.
\item\label{cond1}
The Fisher information matrix (FIM),
	\beqna
	\label{JJJdef}
		\Jmat(\thetavec)\define {\rm{E}}_{\thetavecsmall}
		\left[\nabla_{\thetavecsmall}^T\log f(\xvec;\thetavec)
	\nabla_{\thetavecsmall}\log f(\xvec;\thetavec)\right],
	\eeqna
 is well defined and finite.
\end{enumerate} 
\renewcommand{\theenumi}{\arabic{enumi}}
Then,
\be \label{CR_bound1}
{\rm{E}}_{\thetavecsmall}[\Cmat(\hat{\thetavec},\thetavec)]\succeq \Bmat(\thetavec),~\forall \thetavec\in\Omega_\thetavecsmall,
\ee
where
\be \label{CR_bound2}
\Bmat(\thetavec)\define 	\Lambdamat^{\frac{1}{2}} \Vmat^T\Umat(\Umat^T\Jmat(\thetavec)\Umat)^\dagger
	\Umat^T\Vmat \Lambdamat^{\frac{1}{2}}.
\ee
Equality in \eqref{CR_bound1} is obtained {\em{iff}} the estimation error in the graph-frequency domain satisfies
\beqna \label{equality_cond}
\Lambdamat^{\frac{1}{2}} (\hat{\tilde{\thetavec}}-\tilde{\thetavec})=\Lambdamat^{\frac{1}{2}} \Vmat^T \Umat\left(\Umat^T\Jmat(\thetavec)\Umat\right)^\dagger\Umat^T \nabla_{\thetavecsmall}^T\log f(\xvec;\thetavec),
\eeqna
$\forall \thetavec\in\Omega_\thetavecsmall$.
\end{Theorem}
	\begin{IEEEproof} 
The proof is given in Appendix \ref{App_T3}.
\end{IEEEproof} 
Similar to the explanations in Subsection \ref{Dif_Subs},
it can be seen that the first row and column of the proposed matrix bound  in \eqref{CR_bound2}, $\Bmat({\thetavec})$,  is zero.
Thus, 
in practice, we use the submatrix bound,
 $\Bmat_{\mathcal{M}\setminus 1}({\thetavec})$.
 Discussion of pseudo-inverse matrix bounds  in the general case can be found, for example, in \cite{stoica2001parameter}.

\subsubsection{Relation with the constrained CRB (CCRB) on the MSE}
The CCRB, which was originally presented in \cite{Hero_constraint},  is a
lower bound on the MSE of $\chi$-unbiased estimators   \cite{Eyal_constraint,sparse_con} in constrained parameter estimation settings.
The CCRB 
is especially suitable for the
 case of linear constraints  \cite{Eyal_constraint} and it is attained asymptotically by
the commonly-used constrained maximum likelihood (CML)
estimator \cite{Moore_Sadler_Kozick_2008}.
It can be verified that the proposed bound in \eqref{CR_bound2} is a weighted version of the CCRB, i.e.
\be \label{CR_bound2_2}
\Bmat(\thetavec)=	\Lambdamat^{\frac{1}{2}} \Vmat^T\Bmat_{\text{CCRB}}(\thetavec)\Vmat \Lambdamat^{\frac{1}{2}},
\ee
where the CCRB  is given by \cite{Stoica_Ng}
\beqna
\label{ccrb}
\Bmat_{\text{CCRB}}(\thetavec)=
\Umat(\Umat^T\Jmat(\thetavec)\Umat)^\dagger\Umat^T.
	\eeqna
	This result stems from the fact that the performance criterion in this paper is a weighted version of the MSE with the structure described in \eqref{MSSE}.
The equality condition of the CCRB holds  {\em{iff}}  \cite{Stoica_Ng,Hero_constraint,Eyal_constraint,sparse_con}
\beqna \label{equality_cond_CCRB}
\hat{\thetavec}-\thetavec= \Umat\left(\Umat^T\Jmat(\thetavec)\Umat\right)^\dagger\Umat^T \nabla_{\thetavecsmall}^T\log f(\xvec;\thetavec).
\eeqna
It can be verified that if there is a constrained-efficient estimator which satisfies \eqref{equality_cond_CCRB}, then it is also an estimator which satisfies \eqref{equality_cond}, but not vice versa. The equality conditions of the graph signal recovery are less restrictive, since the error w.r.t. the zero-graph frequency can be neglected.

\subsubsection{Graph CRB on the Dirichlet energy}
The WMSE  bound in Theorem \ref{T3} is a matrix bound. As such, it implies bounds on the marginal WMSE of each node and on submatrices that are related to subgraphs.
In particular,  by applying the trace operator on the bound from \eqref{CR_bound1}-\eqref{CR_bound2}
and using the trace operator properties, \eqref{SVD_new_eq}, 
and \eqref{aSSE}, 
we obtain the associated graph CRB on the expected Dirichlet energy:
		\beqna
		\label{sCRBtrace}
		{\rm{E}}_\thetavecsmall[(\hat{\thetavec}-\thetavec)^T\Lmat (\hat{\thetavec}-\thetavec)]
		\geq{\text{Tr}}(\Bmat(\thetavec))\hspace{2.5cm}
		\nonumber\\
			=	{\text{Tr}}\left( \Lmat \Umat(\Umat^T\Jmat(\thetavec)\Umat)^\dagger\Umat^T\right).
		\eeqna
For an unconstrained estimation problem, in which $\Umat=\Imat_M$,
the bound in \eqref{sCRBtrace} is reduced to 
	\beqna
		\label{sCRBtrace2}			{\rm{E}}_\thetavecsmall[(\hat{\thetavec}-\thetavec)^T\Lmat (\hat{\thetavec}-\thetavec)]\geq {\text{Tr}}(\Bmat(\thetavec))
			=	{\text{Tr}}\left( \Lmat \Jmat^\dagger(\thetavec)\right).
		\eeqna

\subsubsection{Efficiency}
\begin{definition}
\label{def_eff}
A graph-unbiased estimator, in the sense of Proposition \ref{propLehm}, 
that  achieves the graph CRB in Theorem \ref{T3} is said to be an efficient estimator on the graph ${\mathcal{G}}({\mathcal{M}},\xi,\Wmat)$.
    \end{definition}
Similar to the conventional theory on estimators' efficiency, it can be shown that if there is an estimator which satisfies \eqref{equality_cond} and it is not a function of $\thetavec$, then this is an efficient estimator.
 Moreover, similar to the uniformly minimum variance unbiased  estimator \cite[p.~20]{Kayestimation}, the graph-efficient  estimator from Definition \ref{def_eff} is also  the uniformly minimum risk unbiased estimator, i.e. an estimator that is uniformly graph-unbiased (i.e. it satisfies \eqref{unbiased_cond}), and achieves minimum WMSE, defined in \eqref{MSSE}.
From the definition of $\Lmat$,  it can be verified that  $\Lmat^T\yvec =\zerovec $ {\em{iff}} $\yvec = c\onevec_M$, where $c$ is an arbitrary scalar.
Thus,
 if $\hat{\thetavec}_e$ is an  efficient estimator, then  any shifted estimator 
$\hat{\thetavec}_e+c\onevec$, where $c\in{\mathbb{R}}$ is a constant, is also an efficient estimator, since it also satisfies \eqref{equality_cond} and 
is not a function of $\thetavec$.

In the following, the  graph CRB,  estimation methods, and sensor allocation based on the  graph CRB  are developed for the special cases:
	1) linear Gaussian model  with relative measurements (Section \ref{ex1});
	and
2)   recovery of a bandlimited graph signal from corrupted measurements (Section \ref{Bandlim_subsection}).

\section{Example 1: Linear Gaussian model with relative measurements}
\label{ex1}
In this section, we discuss the 
problem of estimating vector-valued node variables
from noisy relative measurements.
This problem arises in many network
applications, such as localization
in sensor networks and motion consensus \cite{Barooah_Hespanha_2007,Barooah_Hespanha_2008},
 synchronization of translations
\cite{Boumal_2013,howard2010estimation}, 
edge flows \cite{Jia_Benson_2019},
and state estimation in power systems \cite{Giannakis_Wollenberg2013,drayer2018detection,GlobalSIP_Drayer_Routtenberg,Grotas_YI_Routtenberg},
where the vector
$\thetavec$  represents parameters such as
 positions, states,  opinions, and voltages. 
 The measurement sensor network in this model (i.e. the measurement graph) may be different from 
 the physical network used in   the   WMSE cost in Subsection \ref{cost_sec}.
 This model represents the fact that 
 many cyber-physical systems consist of two interacting networks: 
 an
underlying {\em{physical}} system with topological structure and a {\em{sensor}}  network topology
that may be with a different topology.
Similarly, 
in other real-life applications
 such as  in genetics, there exist dual networks, with a {\em{physical}}-world network and a second  network that represents the {\em{conceptual or statistical}} world \cite{prabhu2012ultrafast}. 
Another example is in communication systems where
 the topology (local data passage) of internode communication may be different from the  graphical model 
use to describe local dependency \cite{Wiesel_Hero_2012}.
In general, the two topologies do not necessarily   match. In  this section, we use ``bar" to denote the topology associated with the measurements and  determined  by  the  sensing  approach in order to distinguish it from  the  topology  used  in  the  WMSE  (``physical" topology  or a topology that was generated from {\em a-priori} data), described above. 
Finally, in this section we consider a linear model.
 For completeness, the interested reader is referred to an example of the recovery of a nonlinear model with relative measurements in power systems that is described in  \cite{Ariel_Yonina_GlobalSIP}.

\subsection{Model}
\label{model_subsec}
We consider  a noisy measurement of the weighted relative state of each edge, as follows:
\be
\label{model_Gas} 
h_{m,k} = \bar{w}_{m,k}(\theta_m-\theta_k)+\nu_{m,k},~~~\forall (m,k)\in \bar{\xi},
\ee
 where  $\{\nu_{m,k}\}$, $\forall (m,k)\in \bar{\xi}$, $m>k$, is an  i.i.d. Gaussian noise sequence with variance $\sigma^2$.
 The sequence $\{\bar{w}_{m,k}\}$, $\forall (m,k)\in \bar{\xi}$, contains positive weights that are  given by the system parameters
and is assumed to be known.
 We assume that the edge weights satisfy $\bar{w}_{k,m}=\bar{w}_{m,k}$. Thus, from symmetry,
 the measurement and the  noise sequences satisfy
 $h_{k,m}=-h_{m,k}$ and $\nu_{k,m}=-\nu_{m,k}$, respectively, $\forall (m,k)\in \bar{\xi}$.
 In general,
 $\bar{\xi}$ and $\{\bar{w}_{m,k}\}$, that are associated with the measurements and determined by the sensing approach,
 are different from  $\xi$ and $\{{w}_{m,k}\}$, that are based on the physical graph.
 The goal here is to estimate the state vector, $\thetavec=[\theta_1,\ldots,\theta_M]^T$, from the observations in \eqref{model_Gas}. 
 For example, 
 this problem
 with the model in \eqref{model_Gas} where $w_{m,k}=1$,  $\forall (m,k)\in \bar{\xi}$,
 is  the problem of synchronization of translations from
\cite{Boumal_2013,howard2010estimation} .
Another example is PSSE in
electrical networks,  \cite{Giannakis_Wollenberg2013,drayer2018detection,Grotas_YI_Routtenberg}, as described in Section \ref{simulation_sec}.

Let ${\mathcal{G}}({\mathcal{M}},\bar{\xi},\bar{\Wmat})$
be defined as the 
measurement graph associated with the  model in \eqref{model_Gas} over the same vertex set  as in the ``physical" graph, ${\mathcal{G}}({\mathcal{M}},\xi,\Wmat)$, which is used in  \eqref{MSSE}.
We assume that  ${\mathcal{G}}({\mathcal{M}},\bar{\xi},\bar{\Wmat})$ is a connected simple graph and define its associated  Laplacian matrix,   $\bar{\Lmat}{\in{\mathbb{R}}^{M\times M}}$, with the elements
\be
\label{tildeL}
\bar{\Lmat}_{m,k} = 
\left\{
\begin{array}{lr}
  \sum_{k\in \bar{\mathcal{N}}_m}\bar{w}_{m,k}   &  {\text{if }} m=k\\
 -\bar{w}_{m,k}    & {\text{if }} (m,k)\in\bar{\xi}\\
  0   &{\text{otherwise}}
\end{array}
\right.,
\ee
where, similar to Definition \ref{neighborhood},
$\bar{\mathcal{N}}_m = \{k \in {\mathcal{M}} : (m,k) \in \bar{\xi}\}$. In addition, let
the matrix  $\bar{\Emat}$  be the  ${M\times |\bar{\xi}|}$ oriented incidence matrix of the  graph ${\mathcal{G}}({\mathcal{M}},\bar{\xi},\bar{\Wmat})$.  Thus,  each of the columns of $\bar{\Emat}$
has two nonzero elements, $1$ in the $m$th row and $-1$ in the $k$th row, representing an edge  connecting nodes $m$ and $k$, where the sign is chosen arbitrarily  \cite{lu2017closed}.
Then, the model in \eqref{model_Gas} can be rewritten in a matrix form as follows:
\be
\label{model_Gas2} 
\xvec = {\text{diag}}(\bar{\wvec})\bar{\Emat}^T\thetavec+\nuvec,
\ee
where $\xvec$, $\bar{\wvec}$, and $\nuvec$
include the elements $\{h_{m,k}\}$, $\{\bar{w}_{m,k}\}$, $\{\nu_{m,k}\}$,
 respectively, $\forall (m,k)\in \bar{\xi}$, $m>k$, in the same order,
and we used the
symmetry in the model, i.e. the fact that $\bar{w}_{k,m}=\bar{w}_{m,k}$,
 $h_{k,m}=-h_{m,k}$, and  $\nu_{k,m}=-\nu_{m,k}$. By multiplying \eqref{model_Gas2} by $\bar{\Emat}$ from the left  and using
 the fact that \cite{lu2017closed}
 \be
 \label{L_EDE}
 \bar{\Lmat}=\bar{\Emat}{\text{diag}}(\bar{\wvec})\bar{\Emat}^T,
 \ee
we obtain
\be
\label{model_Gas3} 
\bar{\Emat}\xvec = \bar{\Lmat}\thetavec+\bar{\Emat}\nuvec.
\ee
Since $\{\nu_{m,k}\}$,  $m>k$, is an  i.i.d. Gaussian noise sequence with variance $\sigma^2$, then $\bar{\Emat}\nuvec$ is a zero-mean Gaussian vector with a covariance matrix $\sigma^2\bar{\Emat}\bar{\Emat}^T$.

\subsection{Graph CRB and an efficient estimator}
The log-likelihood function for the 
modified model in \eqref{model_Gas3}, after removing constant terms and by using $\bar{\Lmat}^T=\bar{\Lmat}$,  is
\beqna
\label{log_trans}
\log f(\xvec;\thetavec)
=-\frac{1}{2\sigma^2}
(\bar{\Emat}\xvec-\bar{\Lmat}\thetavec)^T(\bar{\Emat}\bar{\Emat}^T)^{\dagger} (\bar{\Emat}\xvec-\bar{\Lmat}\thetavec).\hspace{0.6cm}
\eeqna
Thus, the  gradient of \eqref{log_trans} satisfies
	\beqna
	\label{first_der}
		\nabla_{\thetavecsmall}^T\log f(\xvec;\thetavec)
		=\frac{1}{\sigma^2}\bar{\Lmat}(\bar{\Emat}\bar{\Emat}^T)^{\dagger}\left( \bar{\Emat}{\xvec}-\bar{\Lmat}\thetavec\right).
		\eeqna
		The matrix $\bar{\Emat}\bar{\Emat}^T$ is a Laplacian matrix with $M$ nodes and equal unit weights for all edges. Thus, its
 pseudo-inverse   is given by
\cite{GUTMAN_Xiao_2004}
\be
\label{Einv}
(\bar{\Emat}\bar{\Emat}^T)^{\dagger}=
\left(\bar{\Emat}\bar{\Emat}^T-\frac{1}{M}\onevec\onevec^T\right)^{-1}+\frac{1}{M}\onevec\onevec^T.
\ee	
By substituting \eqref{Einv} in \eqref{first_der}
and using the null-space property, $\bar{\Lmat}\onevec=\zerovec$, one obtains
	\beqna
	\label{first_der2}
		\nabla_{\thetavecsmall}^T\log f(\xvec;\thetavec)
		=\frac{1}{\sigma^2}\bar{\Lmat}
		\left(\bar{\Emat}\bar{\Emat}^T-\frac{1}{M}\onevec\onevec^T\right)^{-1}\left( \bar{\Emat}{\xvec}-\bar{\Lmat}\thetavec\right).
		\eeqna
	Thus,
			the FIM from \eqref{JJJdef} for this case is given by
	\beqna
	\label{J_trans}
	\Jmat(\thetavec)&=&	\frac{1}{\sigma^2}\bar{\Lmat}\left(\bar{\Emat}\bar{\Emat}^T-\frac{1}{M}\onevec\onevec^T\right)^{-1}\bar{\Lmat}.
	\eeqna
	The FIM in \eqref{J_trans} is a function of the graph topology via the Laplacian and the oriented incidence matrices.
	For the special case of unit weights, i.e. for $\bar{\Lmat}=\bar{\Emat}\bar{\Emat}^T$, 
	the FIM satisfies $\Jmat(\thetavec)=	\frac{1}{\sigma^2}\bar{\Lmat}$, which coincides with the result in  \cite{Boumal_2013}.

It is shown in Appendix \ref{pinv_app} that the pseudo-inverse of the FIM in \eqref{J_trans} is given by
	\beqna
	\label{J_trans_inv4}
	\Jmat^\dagger(\thetavec)=		\sigma^2\bar{\Lmat}^\dagger\bar{\Emat}\bar{\Emat}^T
	\bar{\Lmat}^\dagger.
		\eeqna
By substituting \eqref{J_trans_inv4} and $\Umat=\Imat_M$ (since there are no constraints in the considered model) in \eqref{CR_bound1}-\eqref{CR_bound2}, we obtain that the graph CRB for this case is
	\beqna \label{CR_bound2_trans}
	{\rm{E}}_{\thetavecsmall}[\Cmat(\hat{\thetavec},\thetavec)]\succeq\Bmat(\thetavec)=
	\sigma^2\Lambdamat^{\frac{1}{2}} \Vmat^T\bar{\Lmat}^\dagger
	\bar{\Emat}\bar{\Emat}^T
	\bar{\Lmat}^\dagger\Vmat \Lambdamat^{\frac{1}{2}},
\eeqna
which is not a function of the specific values of $\thetavec$.
By applying the trace operator on \eqref{CR_bound2_trans}, we obtain 
the bound on the expected Dirichlet energy, which is
\beqna \label{CR_bound2_trans_trace}
	{\rm{E}}_\thetavecsmall[(\hat{\thetavec}-\thetavec)^T\Lmat (\hat{\thetavec}-\thetavec)]\geq {\text{Tr}}(\Bmat(\thetavec))= \sigma^2{\text{Tr}}(	\bar{\Emat}^T\bar{\Lmat}^\dagger\Lmat\bar{\Lmat}^\dagger\bar{\Emat}).
\eeqna
The graph CRBs are lower bounds on the WMSE and on the Dirichlet energy of any graph unbiased estimators, as defined  for this unconstrained setting in \eqref{unbiased_cond_w_out} in Special case \ref{special_case1}.

For the estimator, by substituting \eqref{first_der},  \eqref{J_trans_inv4}, and $\Umat=\Imat_M$ in \eqref{equality_cond}, equality in \eqref{CR_bound2_trans} is obtained {\em{iff}}
\beqna \label{equality_cond_trans}
\Lambdamat^{\frac{1}{2}} \Vmat^T(\hat{\thetavec}-\thetavec)
=\Lambdamat^{\frac{1}{2}} \Vmat^T \bar{\Lmat}^\dagger \bar{\Emat}^T\bar{\Emat}\bar{\Lmat}^\dagger \bar{\Lmat}  (\bar{\Emat}\bar{\Emat}^T)^{\dagger} (\bar{\Emat}{\xvec}
-\bar{\Lmat}\thetavec).
\eeqna
By using the properties of the Laplacian and incidence matrices, as well as its pseudo-inverse properties, it can be shown that
$\bar{\Emat}^T\bar{\Emat}\bar{\Lmat}^\dagger \bar{\Lmat}  (\bar{\Emat}\bar{\Emat}^T)^{\dagger} \bar{\Emat}=\bar{\Emat}$
and $\bar{\Emat}^T\bar{\Emat}\bar{\Lmat}^\dagger \bar{\Lmat}  (\bar{\Emat}\bar{\Emat}^T)^{\dagger} \bar{\Lmat}=\bar{\Lmat}$.
Thus, 
\eqref{equality_cond_trans} implies the following condition for achievablity of the bound:
\beqna \label{equality_cond_trans2_new}
\Lambdamat^{\frac{1}{2}} \Vmat^T\hat{\thetavec}
&=&\Lambdamat^{\frac{1}{2}} \Vmat^T \bar{\Lmat}^\dagger\bar{\Emat}
{\xvec}
+\Lambdamat^{\frac{1}{2}} \Vmat^T(\Imat_M- \bar{\Lmat}^\dagger\bar{\Lmat})\thetavec 
\nonumber\\
&=&\Lambdamat^{\frac{1}{2}} \Vmat^T \bar{\Lmat}^\dagger \bar{\Emat}
{\xvec}
+\frac{1}{M}\Lambdamat^{\frac{1}{2}} {\Vmat}^T\onevec\onevec^T\thetavec
\nonumber\\
&=&\Lambdamat^{\frac{1}{2}} \Vmat^T \bar{\Lmat}^\dagger  \bar{\Emat}
{\xvec},
\eeqna
where 
the  second  equality is obtained 
by using the fact that for connected graphs there is only one zero eigenvalue of the Laplacian matrix, which is associated with the
eigenvector $\vvec_1=\frac{1}{\sqrt{M}}\onevec$,
and the last equality is obtained  by using the Laplacian null-space property, $\Lambdamat^{\frac{1}{2}}{\Vmat}^T\onevec=\zerovec$.
Since the estimator on the r.h.s. of \eqref{equality_cond_trans2_new} is not a function of  $\thetavec$,
then it
is also an efficient estimator, in the sense of Definition \ref{def_eff}.
By using the null-space property ${\Vmat}^T\onevec=\zerovec$, it can be verified that
any estimator of the form 
\be
\label{est_trans}
\hat{\thetavec}=\bar{\Lmat}^\dagger\bar{\Emat}\xvec +c\onevec,
\ee
 with an arbitrary scalar $c\in{\mathbb{R}}$,
 satisfies the equality condition in \eqref{equality_cond_trans2_new}. Thus, 
the efficient estimator is not unique and the 
true signals, $\thetavec$, can be recovered up to a constant vector, $c\onevec$.
This result is reasonable, since, 
	with relative measurements alone, as given in the model in \eqref{model_Gas}, determining the signal is
possible only up to an additive constant \cite{Barooah_Hespanha_2007,Barooah_Hespanha_2008,Ariel_Yonina_GlobalSIP}.

\subsection{Sample allocation}
\label{opt_sec}
In this subsection, we design a sample allocation rule
for the sensing model from Subsection \ref{model_subsec} based on solving an optimization problem 
that aims to minimize the graph CRB in \eqref{CR_bound2_trans_trace}.
In this case, the graph CRB is achievable  by the efficient estimator in \eqref{est_trans} and is not a function of the unknown graph signal, $\thetavec$. Thus,
minimizing  the graph CRB in \eqref{CR_bound2_trans_trace} will result in the minimum WMSE   over graph unbiased estimators.

In many cases, the physical topology is known and
is the one that is both used  in the cost function in  \eqref{MSSE}
and governs the measurement model  in \eqref{model_Gas} for any edge that is measured. Thus, 
in the following we assume that
the edge weights in \eqref{model_Gas} satisfy
$\bar{w}_{m,k}={w}_{m,k}$, $\forall (m,k)\in \xi \bigcap\bar{\xi}$,
and the sensors can be located only
 in existing edges of the physical  network, where
 if $(m,k)\in\bar{\xi}$, then  $(k,m)\in\bar{\xi}$.
We assume a constrained amount of sensing resources,
e.g. due to limited energy and communication budget.
We thus state the sensor placement problem as follows:
\begin{problem}
\label{prob_1}
 Given a graph ${\mathcal{G}}({\mathcal{M}},\xi,\Wmat)$,
find the smallest subset $\bar{\xi} \subset {\xi}$ such that the graph signal, $\thetavec$, 
  can be correctly recovered (up-to-a-constant) by the 
  measurements on $\bar{\xi}$ and the  graph CRB from \eqref{CR_bound2_trans_trace}, $
 \sigma^2{\text{Tr}}(	\bar{\Emat}^T\bar{\Lmat}^\dagger\Lmat\bar{\Lmat}^\dagger\bar{\Emat})$,  is minimized, where $\bar{w}_{m,k}={w}_{m,k}$, $\forall (m,k)\in \xi \bigcap\bar{\xi}$.
\end{problem}

  The requirement of  correct graph signal recovery in Problem \ref{prob_1} is defined as  recovery up to a constant vector, $c\onevec$, which is an inherent ambiguity with relative measurements.
In order to correctly reconstruct the full graph signal by the estimator from \eqref{est_trans} up-to-a-constant, we need to have ${\text{rank}}(\bar{\Lmat})=M-1$. That is,   for each node, $m\in M$, we need to have at least one edge in $\bar{\xi}$ which is associated with this node. 
It can be shown that this condition for unique up-to-a-constant
recovery is that $\bar{\xi}$ is a spanning tree  of ${\mathcal{G}}({\mathcal{M}},\xi,\Wmat)$ \cite{narasimhan1999data}.
Thus, Problem \ref{prob_1} is equivalent to the following optimization problem.
\begin{problem}
\label{prob_1_re}
 Given a graph ${\mathcal{G}}({\mathcal{M}},\xi,\Wmat)$,
 find the Laplacian matrix $\bar{\Lmat}^*$
such that 
  \beqna
\label{min_sparse}
\bar{\Lmat}^*=
\arg\min_{\bar{\Lmat}}{\text{Tr}}(\bar{\Lmat}^\dagger\bar{\Emat}	\bar{\Emat}^T\bar{\Lmat}^\dagger\Lmat),~~{\text{such that }}\bar{\Lmat}\in S_T({\mathcal{G}}),
\eeqna
where $S_T({\mathcal{G}})$ is the set of 
spanning trees  of ${\mathcal{G}}({\mathcal{M}},\xi,\Wmat)$.
\end{problem}

If $\bar{\Lmat}$ is a spanning tree of a connected (loopless) graph, then it can be seen that its oriented incidence matrix,
$\bar{\Emat}$,
has $M-1$ linearly independent columns.
Thus,  $\bar{\Emat}^T(\bar{\Emat}	\bar{\Emat}^T)^\dagger\bar{\Emat}=\bar{\Emat}^\dagger\bar{\Emat}=\Imat_{M-1}$ and by substituting  \eqref{L_EDE}, we obtain that for
spanning trees
\beqna
\label{bar_bar}
\bar{\bar{\Lmat}}\define \bar{\Lmat}(\bar{\Emat}	\bar{\Emat}^T)^\dagger\bar{\Lmat}
=\bar{\Emat}{\text{diag}}(\bar{\wvec})\bar{\Emat}^T(\bar{\Emat}	\bar{\Emat}^T)^\dagger\bar{\Emat}{\text{diag}}(\bar{\wvec})\bar{\Emat}^T
\nonumber\\
=\bar{\Emat}{\text{diag}}(\bar{\wvec}){\text{diag}}(\bar{\wvec})\bar{\Emat}^T.\hspace{3.9cm}
\eeqna
That is, $\bar{\bar{\Lmat}}$
 is also  a Laplacian matrix of a spanning tree of $\Lmat$ with the weights
$\{{w}_{m,k}^2\}$, $\forall (m,k)\in \bar{\xi}$,
and the optimization from \eqref{min_sparse} is reduced to
\begin{problem}
\label{prob_1_re2}
 Given a graph ${\mathcal{G}}({\mathcal{M}},\xi,\Wmat)$,
 find the Laplacian matrix $\bar{\bar{\Lmat}}^*$
such that 
\beqna
\label{min_sparse2}
\min_{\bar{\bar{\Lmat}}}{\text{Tr}}(\bar{\bar{\Lmat}}^\dagger\Lmat),~~{\text{such that }}\bar{\Lmat}\in S_T({\mathcal{G}}),
\eeqna
where
$\bar{\bar{\Lmat}}=\bar{\Emat}{\text{diag}}(\bar{\wvec}){\text{diag}}(\bar{\wvec})\bar{\Emat}^T$ and ${\bar{\Lmat}}=\bar{\Emat}{\text{diag}}(\bar{\wvec})\bar{\Emat}^T$.
\end{problem}

It is shown in \cite{Spielman_Woo2009}
that 
\be
\label{stretch}
{\text{Tr}}({\bar{\Lmat}}^\dagger\Lmat)=st_{{\bar{\Lmat}}}(\Lmat),
\ee
where $st_{{\bar{\Lmat}}}(\Lmat)$ is the total  stretch of the graph ${\mathcal{G}}({\mathcal{M}},\xi,\Wmat)$ w.r.t. the  spanning tree represented by ${\bar{\Lmat}}$.
Total stretch is a parameter used to measure the quality of a spanning tree in terms of distance preservation
and can represent the average effective resistance 
\cite{Spielman_Woo2009,alon1995graph}.
The objective function in 
Problem \ref{prob_1_re2} is the $\ell_{p=\frac{1}{2}}$ total  stretch  of the graph ${\mathcal{G}}({\mathcal{M}},\xi,\Wmat)$,
i.e.  the  total  stretch from \eqref{stretch} after taking the $p=\frac{1}{2}$
 power of the weights of the edges \cite{cohen2014stretching}. 
Thus, the  problem of finding the minimum graph CRB  is equivalent to
the problem of finding a spanning tree with the minimum average $\ell_{p=\frac{1}{2}}$ stretch, which is a classical problem
in network design, graph theory,  and discrete mathematics.
Although this problem in general is an NP-hard problem,  it was shown that a standard
maximum weight spanning tree algorithm \cite{jungnickel2007graphs} yields
good results in practice 
\cite{Trees_thesis}.
Thus, in this paper we solve the sensor allocation problem in Problem \ref{prob_1_re2} by finding
the maximum weight spanning tree of ${\mathcal{G}}({\mathcal{M}},\xi,\Wmat^2)$,  
which is one of the fundamental  problems of algorithmic graph theory and can be solved by  existing algorithms in near-linear time \cite{jungnickel2007graphs,Neiman2008}.



\section{Example 2: Bandlimited graph signal recovery}
\label{Bandlim_subsection}
In this section, we consider the problem of  signal recovery from noisy, corrupted, and incomplete measurements,  under the constraint of bandlimited graph signals \cite{Chen_Moura_Kovacevic2015,8892652}.
In Subsection \ref{model_bandlimited} we present the model assumed in this section. The graph CRB and Lehmann unbiasedness  are developed in 
Subsection \ref{graph_crb_bandlimited} and sensor  allocation which uses the graph CRB as an objective function is discussed in Subsection  
\ref{opt_sec_bandlimited}.
An efficient estimator for the Gaussian case is presented in Subsection
\ref{eff_bandlimited_sec}.
Finally,  in Subsection \ref{alternative_bandli_sub} we demonstrate the use of 
an alternative cost function, 
which  does not enforce strictly bandlimited constraints.

\subsection{Model}
\label{model_bandlimited}
We assume  the following measurement model:
\be
\label{model1}
\xvec=\Mmat(\thetavec+\wvec),
\ee
where  $\thetavec$ is an unknown signal, $\wvec$ is a noise vector  with
  zero mean  and  a known distribution,
and $\Mmat \in[0,1]^{D\times M}$ is a binary mask matrix with $D\leq M$.
The mask matrix, $\Mmat$, may indicate missing data, i.e. the indicator matrix for the missing values \cite{Yankelevsky_Elad2016}, removing data which is irrelevant for a specific task \cite{Bayram_Frossard_2020,Mao2019}, as well as representing any interpolation, filtering, and normalization approaches \cite{Chen_Moura_Kovacevic2015}.
The task is to recover the true signal, $\thetavec$,
based on the accessible measurement vector,
$\xvec$.
Signal recovery from inaccessible and corrupted measurements requires
additional knowledge of signal properties.
In GSP, a widely-used  assumption is that the signal of interest, $\thetavec$, is a graph-bandlimited signal \cite{Sandryhaila_Moura_2013,Shuman_Ortega_2013,Chen_Kovavic_2016}, i.e. its GFT, $\tilde{\thetavec}$, defined in \eqref{GFT}, is a $R$-sparse 
vector.
Here
we  assume 
  a graph low-frequency signal defined as follows:
\begin{definition}
\label{band_lim}
 A graph signal $\thetavec\in{\mathbb{R}}^M$ is bandlimited w.r.t.
a GFT basis $\Vmat$, as defined in \eqref{GFT}, with bandwidth $R$ when
\be
\label{RRR}
\tilde{\theta}_m=0,~~~  m=R+1,\ldots,M.
\ee
\end{definition}
The condition in \eqref{RRR} 
can be represented as	
the linear constraint from \eqref{linear} with 
\beqna
\label{const1}
\Gmat=\Qmat\Vmat^T~{\text{and }}\avec=\zerovec_{(M-R)\times 1},
\eeqna	where $\Qmat$ is an $(M-R)\times M$ matrix of the form $\
\Qmat=\left[ \zerovec_{(M-R)\times R}, \Imat_{M-R}\right]$.
		Thus, a null-space matrix $\Umat$, defined in \eqref{one}, can be chosen to be
		$\Umat=\Vmat\bar{\Qmat}$, where $\bar{\Qmat}=\left[
\begin{array}{c}\Imat_{R}\\
\zerovec_{(M-R)\times R}\end{array}\right]$.
\subsection{Graph CRB and Lehmann unbiasedness}
\label{graph_crb_bandlimited}
By substituting the model from Subsection \ref{model_bandlimited} in \eqref{unbiased_cond} it can be verified that in this case
an estimator, $\hat{\thetavec}$, is a Lehmann-unbiased estimator of $\thetavec$
	 under the constrained set in \eqref{RRR}
{\em{iff}}
\beqna
\label{unbiased_cond_bl}
\bar{\Qmat}^T\Vmat^T \Lambdamat^{\frac{1}{2}} \Vmat^T 
{\rm{E}}_{\thetavecsmall}[ \hat{\thetavec}-\thetavec]=\zerovec, ~\forall \thetavec\in \Omega_\thetavecsmall^R,
\eeqna
where $\Omega_\thetavecsmall^R$ is the 
subspace of $R$-bandlimited graph signals that satisfy Definition \eqref{band_lim}.
By using the GFT operator from \eqref{GFT}, as well as the fact that $\lambda_1=0$, it can be verified that
the unbiasedness condition in \eqref{unbiased_cond_bl} is equivalent to requiring that 
\be
\label{unbiased_for_band_lim}
{\rm{E}}_{\thetavecsmall}[ 
\hat{\tilde{\theta}}_m-\tilde{\theta}_m]=0,~~~ 2\leq m\leq R.
\ee
The condition in \eqref{unbiased_for_band_lim} reflects the fact that the high graph frequencies are known to be zero for a $R$-bandlimited graph signal, and, thus, there is no need for an unbiasedness condition on  frequencies higher than $R$. 
In addition,
since the proposed cost
is oblivious to the
 estimation error of a constant bias over the graph, $c \onevec$, 
 then the estimator  of the first graph frequency, $\hat{\tilde{\theta}}_1$, can also  have an arbitrary bias.

By substituting $\Umat=\Vmat\bar{\Qmat}$, $\Vmat^T\Vmat=\Imat$, 
and the model from \eqref{model1}
in \eqref{CR_bound2}, we obtain that
the proposed graph CRB
for graph bandlimited signals 
is given by
\be \label{CR_bound2_freq_band}
\Bmat(\thetavec)= 	\Lambdamat^{\frac{1}{2}} \bar{\Qmat}\left(\bar{\Qmat}^T\Vmat^T\Mmat^T\Jmat(\Mmat\thetavec)\Mmat\Vmat\bar{\Qmat}\right)^\dagger \bar{\Qmat}^T \Lambdamat^{\frac{1}{2}},
\ee
where 
according to \eqref{JJJdef}, 
$\Jmat(\Mmat\thetavec)\in {\mathbb{R}}^{D\times D}$ is the  FIM
w.r.t. the unknown parameter vector $\Mmat\thetavec$.
Thus, by a reparametrization  of the FIM,
it can be seen  that $\Jmat(\thetavec)=\Mmat^T \Jmat(\Mmat\thetavec)\Mmat$.

Since $\Lambdamat_{1,1}=0$, and due to the structure of $\bar{\Qmat}$, the first row and the last $M-R$ rows of $\Lambdamat^{\frac{1}{2}} \bar{\Qmat}$ are zero rows. Thus,  the relevant bound in \eqref{CR_bound2_freq_band}
is the submatrix $\Bmat_{\mathcal{M}\setminus 1}(\thetavec)$, which deploys a bound in the estimation error associated with the low graph frequencies. 
As a result, by substituting \eqref{SVD_new_eq} and   \eqref{band_lim} in \eqref{CR_bound2_freq_band},
applying the trace operator on the result and  using its properties, as well as  the definition of $\bar{\Qmat}$, and under the assumption that  $R\leq D$,  we obtain that
the trace  graph CRB is given by
\beqna \label{sCRBtrace_bandlimited}
{\text{Tr}}(\Bmat(\thetavec))=
	\sum_{m=2}^R \lambda_m \left[(\Vmat_{\mathcal{M},\mathcal{R}}^T\Mmat^T\Jmat(\Mmat\thetavec)\Mmat\Vmat_{\mathcal{M},\mathcal{R}})^{-1}\right]_{m,m},
\eeqna
where $\mathcal{R}=\{1,\ldots,R\}$.
According to \eqref{sCRBtrace},  \eqref{sCRBtrace_bandlimited} is a lower bound on the expected Dirichlet energy.
 It can be seen that a necessary and sufficient condition for a unique up-to-a-constant
reconstruction of a graph bandlimited signal  as defined in Definition \eqref{band_lim}, i.e. the reconstruction of $\tilde{\theta}_m$, $m=2,\ldots,R$, is that ${\text{rank}}(\Vmat_{\mathcal{M},\mathcal{R}}^T\Mmat^T\Jmat(\Mmat\thetavec)\Mmat\Vmat_{\mathcal{M},\mathcal{R}})=
R$, which is determined by the properties of the matrix $\Mmat$.

For the special case where  
$\Mmat$ is a sampling matrix associated with the subset of nodes ${\mathcal{S}}$, i.e. it  satisfies 
$\Mmat_{\mathcal{D},{\mathcal{S}}}=\Imat_{{\mathcal{S}}}$,
$\Mmat_{\mathcal{D},{\mathcal{S}}^c}=\zerovec$,
$\mathcal{D}=\{1,\ldots,D\}$,
and $D=|{\mathcal{S}}|$, then, 
according to the model in \eqref{model1},
$\xvec=\thetavec_{\mathcal{S}}+\wvec_{\mathcal{S}}$.
Thus,
$
\Mmat\Vmat_{\mathcal{M},\mathcal{R}}=
\Vmat_{\mathcal{S},\mathcal{R}}
$
and \eqref{sCRBtrace_bandlimited}
is reduced to 
\beqna
\label{zzz}
{\text{Tr}}(\Bmat(\thetavec))= 
	\sum\nolimits_{m=2}^R \lambda_m \left[(\Vmat_{\mathcal{S},\mathcal{R}}^T\Jmat(\thetavec_{\mathcal{S}}) \Vmat_{{\mathcal{S}},\mathcal{R}})^{-1}\right]_{m,m}.
\eeqna
The r.h.s. of \eqref{zzz} demonstrates the fact that while we only have access to information on a subset of the states, $\thetavec_{\mathcal{S}}$, we still have a well-defined estimator of the full vector $\thetavec$ as long as $\Vmat_{\mathcal{S},\mathcal{R}}^T\Jmat(\thetavec_{\mathcal{S}}) \Vmat_{{\mathcal{S}},\mathcal{R}}$ is a non-singular matrix.

Similar to the derivation of \eqref{zzz}, by substituting $\Umat=\Vmat\bar{\Qmat}$, $\Vmat^T\Vmat=\Imat$, and the model from \eqref{model1} with $\Mmat_{\mathcal{D},{\mathcal{S}}}=\Imat_{{\mathcal{S}}}$ and
$\Mmat_{\mathcal{D},{\mathcal{S}}^c}=\zerovec$,
in \eqref{ccrb}, we obtain that
the trace CCRB
for graph bandlimited signals 
is given by
\beqna \label{ccrb_freq_band2}
{\text{Tr}}\left(\Bmat_{\text{CCRB}}(\thetavec)\right)= 	{\text{Tr}}\left( \left(\bar{\Qmat}^T\Vmat^T\Mmat^T\Jmat(\thetavec_{\mathcal{S}})\Mmat\Vmat\bar{\Qmat}\right)^{-1} \right) 
\nonumber\\
=\sum_{m=1}^R \left[(\Vmat_{\mathcal{S},\mathcal{R}}^T\Jmat(\thetavec_{\mathcal{S}})\Vmat_{\mathcal{S},\mathcal{R}})^{-1}\right]_{m,m},
\eeqna
under the assumption that $\Vmat_{\mathcal{S},\mathcal{R}}^T\Jmat(\thetavec_{\mathcal{S}})\Vmat_{\mathcal{S},\mathcal{R}}$ is a non-singular matrix. Thus, the CCRB is not affected by the eigenvalues of the Laplacian matrix (i.e. by the graph frequencies).
For the special case of an i.i.d. Gaussian model, i.e. when $\Jmat(\thetavec)=\frac{1}{\sigma^2}\Imat$, the CCRB in \eqref{ccrb_freq_band2}
satisfies
\beqna \label{ccrb_freq_band3}
{\text{Tr}}\left(\Bmat_{\text{CCRB}}(\thetavec)\right)=\sigma^2 
{\text{Tr}}\left((\Vmat_{\mathcal{S},\mathcal{R}}^T\Vmat_{\mathcal{S},\mathcal{R}})^{-1}\right).
\eeqna
The conventional CCRB on the  r.h.s. of  \eqref{ccrb_freq_band3} has been used 
as an objective function for graph sampling of  bandlimited graph signals in  \cite{anis2016efficient}, where this policy is called A-optimal design (up to a multiplication by $\sigma^2$).


\subsection{Sensor  allocation}
\label{opt_sec_bandlimited}
The graph CRB in \eqref{zzz} allows us to define a criterion for  sampling set selection.
We thus state the sensor placement problem as follows:
\begin{problem}
\label{prob_2}
 Given a graph ${\mathcal{G}}({\mathcal{M}},\xi,\Wmat)$,
 a  cutoff frequency,
$\lambda_R$,
 and a number of sensors $D$, find the  subset  of nodes (sensor placements) $\mathcal{S}^*$ such that all   $R$-bandlimited signals 
can be uniquely recovered from their samples
on this subset and the worst-case graph CRB from \eqref{zzz} is minimized, i.e. 
 \[
 \hspace{-0.2cm}\mathcal{S}^*=\arg\min_{\mathcal{S}:|\mathcal{S}|=D} \max_{\thetavecsmall\in \Omega_\thetavecsc^R}
 \nonumber\\\sum_{m=2}^R \lambda_m \left[(\Vmat_{\mathcal{S},\mathcal{R}}^T\Jmat(\thetavec_{\mathcal{S}})\Vmat_{\mathcal{S},\mathcal{R}})^{-1}\right]_{m,m}.
 \]
\end{problem}
In order for this problem to be well defined, we should consider only subsets such that $\Vmat_{\mathcal{S},\mathcal{R}}^T\Jmat(\thetavec_{\mathcal{S}})\Vmat_{\mathcal{S},\mathcal{R}}$ is a full rank matrix.  In particular, we require  $D\geq R$.
Since finding the optimal set in Problem \ref{prob_2} is an NP-hard problem, 
a greedy algorithm is described in Algorithm \ref{Alg1}.
At each iteration of this algorithm we remove the node
 that maximally increases the graph CRB  in \eqref{zzz} (maximized w.r.t. $\thetavec$), until we obtain the subset $\mathcal{S}\subset {\mathcal{M}}$ with cardinality $D$.
The greedy approach removes one sample at each iteration  rather than adding one sensor at a time in order for the matrix $\Vmat_{\mathcal{L}^{(i)}\setminus w,\mathcal{R}}^T\Jmat(\thetavec_{\mathcal{L}^{(i)}\setminus w})\Vmat_{\mathcal{L}^{(i)}\setminus w,\mathcal{R}}$ from the r.h.s. of \eqref{objective} to be a non-singular matrix at each step.
 The complexity of this greedy approach can be reduced even further by applying methods with  reduced complexity (see, e.g. \cite{Sakiyama_Tanaka_Ortega2019,wang2019low,bai2020fast,ortiz2019sparse}).
\begin{algorithm}[H]
	\algorithmicrequire{1) Laplacian matrix, $\Lmat=\Vmat\Lambdamat \Vmat^T$}, 2) number of sensors $D\geq R$, and 3) cutoff frequency, $R$\\
	\algorithmicensure{ Subset of sensors, $\mathcal{S}$}
	\begin{algorithmic}[1]
		\STATE Initialize the set of available locations, $\mathcal{L}^{(0)}={\mathcal{M}}$
		\STATE Initialize iteration index, $i=0$
		 \WHILE{$|\mathcal{L}^{(i)}|>M-D$}
	 \STATE Find the optimal node to remove:
	 \vspace{-0.15cm}
\beqna
\label{objective}
w
 = \arg \min_{w\in {\mathcal{L}}^{(i)}} \max_{\thetavecsmall\in \Omega_\thetavecsc^R}\sum\nolimits_{m=2}^R \lambda_m \hspace{2.5cm}
 \nonumber\\
 \times \left[(\Vmat_{\mathcal{L}^{(i)}\setminus w,\mathcal{R}}^T\Jmat(\thetavec_{\mathcal{L}^{(i)}\setminus w})\Vmat_{\mathcal{L}^{(i)}\setminus w,\mathcal{R}})^{-1}\right]_{m,m}
 \eeqna
 		 \STATE  Update the available locations,
 		 $\mathcal{L}^{(i)}
		\leftarrow \mathcal{L}^{(i-1)}\setminus w$, and the iteration index, $i
		\leftarrow i+1$
		\ENDWHILE
		\STATE 	Update the set of removed locations: $\mathcal{S}=\mathcal{M}\setminus\mathcal{L}^{(i)}$
	\end{algorithmic}
	\caption{Sensor allocation for bandlimited graph signals}
	\label{Alg1}
\end{algorithm} 
	\vspace{-0.75cm}

\subsection{Efficient estimator for the Gaussian case}
\label{eff_bandlimited_sec}
The CML estimator of $\tilde{\thetavec}$
is obtained by maximizing the log-likelihood function of the model in \eqref{model1} under the constraint in \eqref{RRR}.
Then,
by using the invariance property of the ML estimator \cite{Kayestimation}, the CML estimator of ${\thetavec}$ is given by $
\hat{{\thetavec}}^{\text{CML}}=\Vmat\hat{\tilde{\thetavec}}$.
In the following, we develop an efficient estimator 
for the special case of Gaussian noise,
which is also the CML estimator in this case. Thus, in the following, 
$\wvec$ is  zero-mean Gaussian noise with a known covariance matrix, $\bsigma$,  and, thus, (p. 47, \cite{Kayestimation}) \be\Jmat(\thetavec)=\Mmat^T(\Mmat\bsigma\Mmat^T)^{-1} \Mmat.
\ee
By substituting $\Umat=\Vmat\bar{\Qmat}$ and the model from  \eqref{model1}  in
\eqref{equality_cond} and using $\Vmat^T \Vmat=\Imat$, we obtain
\beqna \label{equality_cond_freq2}
\Lambdamat^{\frac{1}{2}}  (\hat{\tilde{\thetavec}}-\tilde{\thetavec})=
\Lambdamat^{\frac{1}{2}} \bar{\Qmat}\left(\bar{\Qmat}^T	 \Vmat^T\Mmat^T(\Mmat\bsigma\Mmat^T)^{-1} \Mmat\Vmat\bar{\Qmat}\right)^\dagger
\nonumber\\
\times
\bar{\Qmat}^T	\Vmat^T\Mmat^T(\Mmat\bsigma\Mmat^T)^{-1}\left(\xvec- \Mmat\Vmat\tilde{\thetavec}\right),
\eeqna
$\forall \thetavec \in \Omega_\thetavecsmall^R$, where  we use $\Vmat^T\Vmat=\Imat_M$.
Under the assumption that
$\tilde{\thetavec}$ is a 
$R$-bandlimited graph signal as defined in Definition \ref{band_lim} and that ${\text{rank}}( \Mmat\Vmat)=R$,
it can be verified that 
\beqna
\label{est1}
\hat{\tilde{\thetavec}}=
\bar{\Qmat}\left(\bar{\Qmat}^T	 \Vmat^T\Mmat^T(\Mmat\bsigma\Mmat^T)^{-1} \Mmat\Vmat\bar{\Qmat}\right)^\dagger
\nonumber\\
\times\bar{\Qmat}^T	 \Vmat^T\Mmat^T(\Mmat\bsigma\Mmat^T)^{-1}\xvec\hspace{1.5cm}
\eeqna
satisfies the efficiency condition in \eqref{equality_cond_freq2}.
By using the definition of $\bar{\Qmat}$,
the estimator from \eqref{est1} 
 can be written as
\beqna
\label{estfreq}
\hat{\tilde{\thetavec}}_{\mathcal{R}}=\left(\bar{\Qmat}^T	\Vmat^T\Mmat^T(\Mmat\bsigma\Mmat^T)^{-1} \Mmat\Vmat\bar{\Qmat}\right)^\dagger\nonumber\\
\times\bar{\Qmat}^T	\Vmat^T\Mmat^T(\Mmat\bsigma\Mmat^T)^{-1}\xvec\hspace{1cm}
\\
\label{estfreq_zero}
\hat{\tilde{\thetavec}}_{\mathcal{M}\setminus \mathcal{R}}=\zerovec. \hspace{4.25cm}  
\eeqna
Since $\hat{\tilde{\thetavec}}$ is also 
the CML estimator of $\tilde{\thetavec}$, then,
by using the invariance property  and 
\eqref{estfreq}-\eqref{estfreq_zero}, the CML estimator of ${\thetavec}$ is given by $
\hat{{\thetavec}}^{\text{CML}}=\Vmat\hat{\tilde{\thetavec}}$. 
Once the full-rank assumption of $\bar{\Qmat}^T	\Vmat^T\Mmat^T(\Mmat\bsigma\Mmat^T)^{-1} \Mmat\Vmat\bar{\Qmat}$ is
satisfied, we can find a sampling operator to achieve
perfect recovery for a given $M$.

\subsection{Alternative approach for bandlimited graph signal recovery}
\label{alternative_bandli_sub}
In Subsections \ref{model_bandlimited}-\ref{eff_bandlimited_sec} we present the estimation approach that aims to minimize the WMSE from \eqref{MSSE} under the strict constraints of zero high graph frequencies from \eqref{RRR}.
In this subsection we present an alternative approach for bandlimited graph signal recovery  based on minimizing  the WMSE only over the low graph frequencies, and without constraints. That is, we use the expected cost function from \eqref{cost_bandli}
with the frequency band $\mathcal{S}=\{1,\ldots,R\}$, which satisfies
 \beqna
 \label{risk_bandli}
 {\rm{E}}\left[\Lambdamat_{\mathcal{R}}^{\frac{1}{2}}\Vmat_{\mathcal{M},\mathcal{R}}^T (\hat{\thetavec}-\thetavec)(\hat{\thetavec}-\thetavec)^T
 \Vmat_{\mathcal{M},\mathcal{R}}\Lambdamat_{\mathcal{S},\mathcal{R}}^{\frac{1}{2}}\right]
 \nonumber\\
 =\Lambdamat_{\mathcal{R}}^{\frac{1}{2}} {\rm{E}}[ (\hat{\tilde{\thetavec}}-\tilde{\thetavec})(\hat{\tilde{\thetavec}}-\tilde{\thetavec})^T
]\Lambdamat_{\mathcal{S},\mathcal{R}}^{\frac{1}{2}},
 \eeqna
without forcing constraints.
Similar to the derivations of the
graph unbiasedness in Proposition \ref{propLehm},
it can be shown that in this case the Lehmann unbiasedness condition is identical to the condition in \eqref{unbiased_for_band_lim}.
Similar to the derivation of the 
graph CRB from Theorem \ref{T3} in Appendix \ref{App_T3},
for any estimator  $\hat{\thetavec}$  that is an unbiased estimator of $\thetavec$ in the sense of \eqref{unbiased_for_band_lim},
we can obtain the following bound on the risk in \eqref{risk_bandli}:
\be 
\label{bandlimited_bound}
\tilde{\Bmat}(\thetavec)=	\Lambdamat_{\mathcal{R}}^{\frac{1}{2}} \Vmat_{\mathcal{M},\mathcal{S}}^T\Jmat^\dagger(\thetavec)
\Vmat_{\mathcal{M},\mathcal{R}}\Lambdamat^{\frac{1}{2}}.
\ee
By applying the trace operator on \eqref{bandlimited_bound}
and substituting the model from \eqref{model1}, we obtain the same scalar bound as the bound on the WMSE under constraints in \eqref{sCRBtrace_bandlimited}.
Finally,   the alternative cost function from \eqref{risk_bandli} implies that   the low-graph-frequencies should be estimated by the estimator in \eqref{estfreq}.
However, this alternative cost  does not force the estimator to satisfy the bandlimitedness constraints, as in  \eqref{estfreq_zero}.
This approach fits the practical setting when the original signal is only approximately bandlimited \cite{anis2016efficient}.

\section{Simulations}
\label{simulation_sec}
In this section,  we  evaluate the performance
of the proposed graph CRB, estimators, and sensor allocation methods.
In Subsection \ref{Ex1_sim} we present simulations of the linear Gaussian model with relative measurements, as described in  Section \ref{ex1}.
In Subsection \ref{Ex2_sim} we present simulations of bandlimited graph signal recovery,
as described in  Section \ref{Bandlim_subsection}.
The  performance of all  estimators   is evaluated using at least $1,000$ Monte-Carlo simulations.

We consider the following 
 two test cases:
\subsubsection{State estimation in electrical networks}
The PSSE problem is described in Subsection \ref{sub_sec_motivation}.
The simulations of this test case are
implemented  on the IEEE 118-bus system from \cite{iEEEdata} that represents a portion of the American Electric Power System and has $M=118$ vertices. Sensor allocation in power systems is of great importance (see, e.g.  \cite{zhao2014identification} and references therein).
\subsubsection{Graph signal recovery in random graphs}
We simulate synthetic graphs from the Watts-Strogatz small-world graph model \cite{Watts_Strogatz} with  varying numbers of nodes, $M$, and an average nodal degree  of $4$.
In addition, we simulate  the Erd\H{o}s-R$\acute{\text{e}}$nyi graph model, which is constructed
by connecting nodes randomly, where each edge is included
in the graph with a probability that is independent of any other edge.

\subsection{Linear Gaussian model with relative measurements}
\label{Ex1_sim}
In this subsection we consider the linear Gaussian model with relative measurements from Section \ref{opt_sec}.
We
compare
 the estimation performance of the estimator from
 \eqref{est_trans} and the proposed graph CRB from \eqref{CR_bound2_trans_trace} for the following  sensor allocation policies:
 \begin{enumerate}
\item Max. ST - 
 maximum spanning tree of ${\mathcal{G}}({\mathcal{M}},\xi,\Wmat^2)$, which is the approximated solution of Problem \ref{prob_1_re2}.
\item Min. ST - minimum spanning tree of ${\mathcal{G}}({\mathcal{M}},\xi,\Wmat^2)$, which can be considered as the worst-case solution of  Problem \ref{prob_1_re2}.
 \item Rand ST - 
 arbitrary spanning tree of ${\mathcal{G}}({\mathcal{M}},\xi,\Wmat^2)$, choosing randomly over the network.
\end{enumerate}
To find the minimum and maximum spanning trees in 1 and 2 we use the MATLAB function {\em{graphminspantree}}, which has a computational complexity of ${\cal{O}}(M^2
\log(M))$.

 First, we consider PSSE 
 with supervisory control and data acquisition (SCADA)
sensors  that measure
the power flow at the edges.
The
linear approximations of the  power
flow model, named  DC model  \cite{Abur_book}, which
represents  the measurement vector of the active powers at the buses, $\xvec$,
can be written
as the model in 
\eqref{model_Gas}, where
$\theta_m$  is the voltage phase at bus $m$, $ m=1,\ldots,M$, $M=118$, and 
$\bar{w}_{m,k}=w_{m,k}$, $\forall (m,k)\in \bar{\xi}$, are the susceptances of the transmission lines.
Since the DC model  is an up-to-a-constant model,  conventional PSSE considers a reference bus with $\hat{\theta}_1=0$ and then uses the ML estimator  of the other parameters,
$\hat{\thetavec}_{\mathcal{M}\setminus 1}  = \left(\Lmat_{\mathcal{M},\mathcal{M}\setminus 1}\right)^\dagger\bar{\xvec}$.
This procedure is equivalent to the  simulated estimator from
\eqref{est_trans}, where $c$ is chosen such that $\hat{\theta}_1=0$.

\begin{figure}[htb]
     \centering
\subcaptionbox{\label{fig:SNR_fig}}[\linewidth]
{\includegraphics[width=8cm]{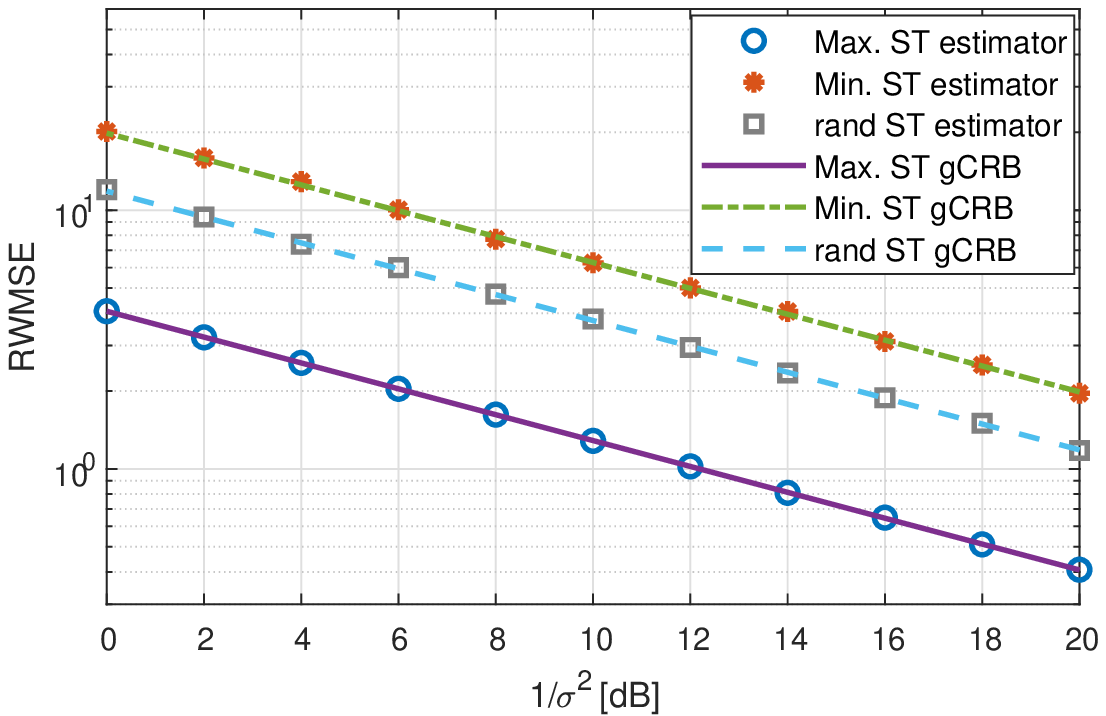}}
\subcaptionbox{\label{fig:rand_Fig}}[\linewidth]
{ \includegraphics[width=8cm]{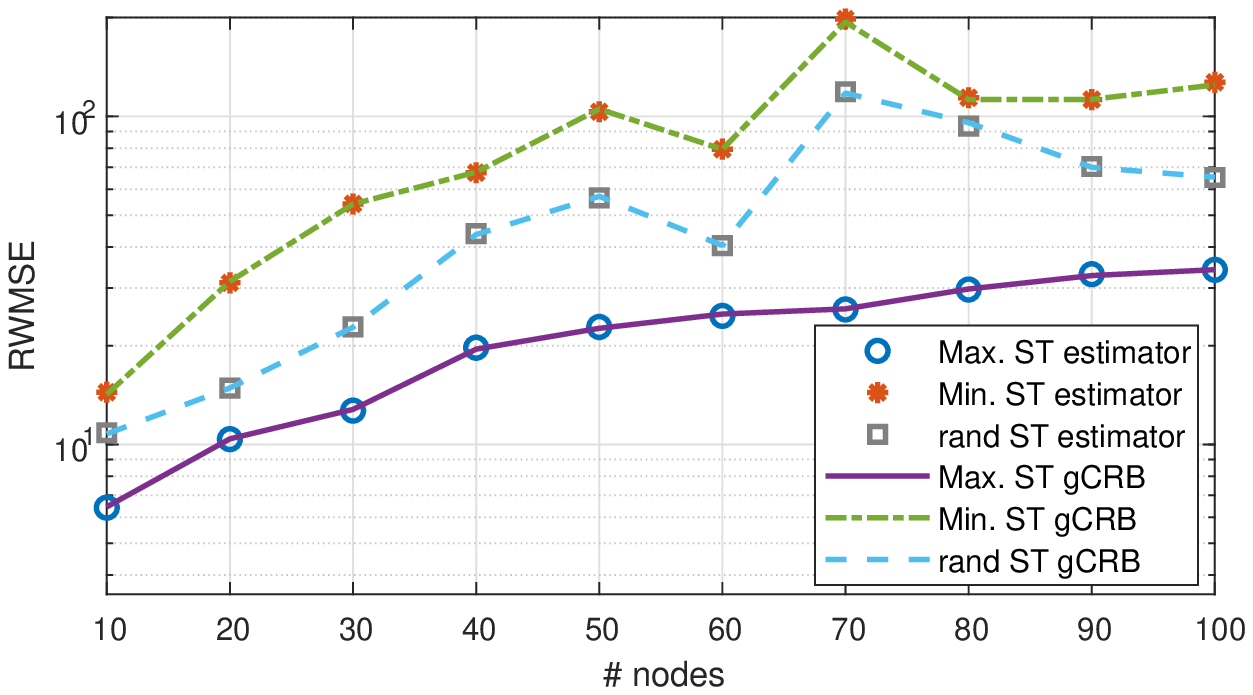}}
\vspace{-0.1cm}
     \caption{Linear Gaussian model with relative measurements: The  graph CRB and the root WMSE  for the three sensor allocation policies (Max. ST, Min. ST, and rand ST)
 for a)  state estimation in power systems  versus $\frac{1}{\sigma^2}$, in IEEE 118-bus system,  $M=118$; and b)
 signals in random graphs versus the number of nodes in the network, for $\sigma^2=1$.}
 \label{Fig1}
 \vspace{-0.25cm}
 \end{figure}

The root WMSE  (square of the Dirichlet energy)
 of the  estimator from  \eqref{est_trans} and the root of the graph CRB are presented  for {\em{PSSE in electrical networks}} under  the different sensor allocation policies are presented in Fig. \ref{Fig1}.a
 versus  $\frac{1}{\sigma^2}$, where $\sigma^2$ is the variance of the
   Gaussian noise, $ \nu_{m,k}$, from \eqref{model_Gas}.
Similarly, 
in Fig. \ref{Fig1}.b the  estimation performance and the root of the graph CRB are presented for a {\em{random graph}}
 with the  three sample allocation policies versus the number of nodes in the system for $\sigma^2=1$.
  It can be seen that in both figures the maximum spanning tree policy  has significantly lower Dirichlet energy than that of the 
   random and the minimum spanning tree schemes, where the minimum spanning tree is worse than   random sampling by an arbitrary spanning tree.
The results indicate  that by applying knowledge of the physical nature of the grid, we can achieve
significant performance gain by using a limited number of well-placed sensors.
Finally, it can be verified that since the estimator from  \eqref{est_trans} is an efficient estimator  on the graph ${\mathcal{G}}({\mathcal{M}},\xi,\Wmat)$ in the sense of Definition \ref{def_eff}, then it coincides with the associated graph CRB for all the considered scenarios. 
 \vspace{-0.25cm}
\subsection{Bandlimited graph signal recovery}
\label{Ex2_sim}
In this subsection we consider the problem of 
bandlimited graph signal recovery  from Section
\ref{Bandlim_subsection} for electrical networks, where $\wvec$ is  zero-mean Gaussian noise with a known diagonal covariance matrix.
The noise variance is set to $\sigma_l^2$ for the 64 buses (vertices) with loads and $\sigma_g^2=0.5 \sigma_l^2$ for the 54 buses with generators.
Thus, in this case
$\Jmat(\thetavec)$ is a diagonal matrix with 
twos and fours on its diagonal, associated with load and generator buses, respectively. The signal, $\thetavec$, is set to the values in \cite{iEEEdata}.
 
We assume a PSSE with direct access to  a limited number of state (voltage) measurements. Thus, we assume the model in \eqref{model1} for the special case where  
$\Mmat$ is a sampling matrix associated with the subset of nodes ${\mathcal{S}}$, i.e. it  satisfies 
$\Mmat_{\mathcal{D},{\mathcal{S}}}=\Imat_{{\mathcal{S}}}$ and
$\Mmat_{\mathcal{D},{\mathcal{S}}^c}=\zerovec$. This can be obtained in practical electrical networks by using Phasor Measurement Units (PMUs)   \cite{PMUbook}.
However,
 installing PMUs on all possible buses is impossible
 due to budget constraints and limitations
 on power and communication resources.  Thus, there is a need to establish a method
to determine which information should be observed
in the course of designing electrical networks.
In power systems, the voltage signal,  $\thetavec$,  is shown to be smooth  \cite{GlobalSIP_Drayer_Routtenberg,drayer2018detection}.
 However, the existing PSSE methods do not incorporate the smoothness and bandlimitedness constraints.
 
We compare the estimation performance of the CML estimator from  \eqref{estfreq}-\eqref{estfreq_zero} and the proposed graph CRB from \eqref{zzz}  
for the following sensor selection policies:
\begin{enumerate}
    \item Random sensor allocation policy (rand.) - named also uniform sampling \cite{Chen_Kovavic_2016}, in which sample indices are chosen
from ${1,\ldots,M}$ independently and randomly, where we bound the condition number of the matrix $(\bar{\Qmat}^T	 \Vmat^T\Mmat^T
\Mmat\Vmat\bar{\Qmat})$ of the chosen set. 
    \item  Minimum graph CRB (Alg. 1) - the sensor allocation policy from Algorithm \ref{Alg1}.
\item Experimentally designed
sampling (E-design) \cite{chen2015discrete} - that maximizes the smallest singular
value of the matrix  $\Mmat \Vmat_{\mathcal{M},\mathcal{R}}=\Vmat_{{\mathcal{S}},\mathcal{R}}$.
\item A-optimal design  (A-design) \cite{anis2016efficient}  - which is equivalent to minimizing the CCRB under i.i.d. Gaussian assumptions 
on the r.h.s. of \eqref{ccrb_freq_band3}.
\end{enumerate}
The objective functions in 2)-4)  lead to combinatorial problems, and, thus, are implemented here by greedy  algorithms. In order to have a fair comparison, we implement all methods by
Algorithm \ref{Alg1}, where \eqref{objective} is replaced by the chosen objective.

\begin{figure}[htb]
\subcaptionbox{\label{fig:SNR_2}}[\linewidth]
{\includegraphics[width=7.5cm]{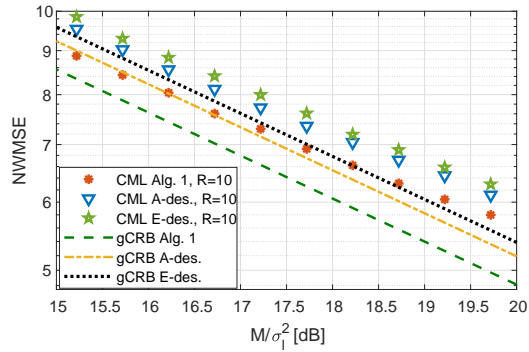}}
\subcaptionbox{\label{fig:rand_Fig2}}[\linewidth]
{ \includegraphics[width=7.5cm]{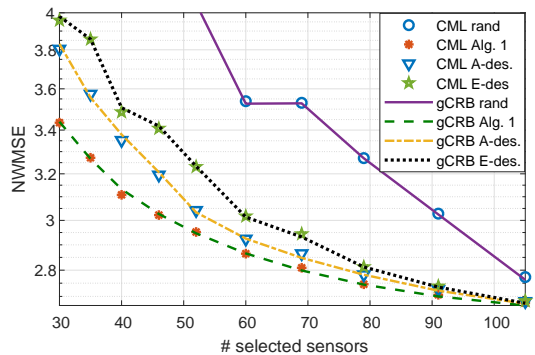}}
     \caption{Bandlimited graph signal recovery: The graph CRB and the root WMSE  for a random sensor allocation policy, and for  Algorithm \ref{Alg1}, A-design, and E-design methods   for state estimation in power systems  in IEEE 118-bus system  (with $M=118$) for a) real-data values and assumed $R=10$ cutoff frequency and $D=40$ sensors versus   $\frac{M}{\sigma_l^2}$; and b) $R=10$ bandlimited graph signal versus $D$
     for $\sigma_l^2=0.5$.}
     \label{Fig2}
\end{figure}

 In Figs. \ref{Fig2}.a and \ref{Fig2}.b the  root WMSE  
 of the CML estimator from  \eqref{estfreq}-\eqref{estfreq_zero} and the root of the graph CRB are presented for the IEEE 118-bus test case 
 with the sampling policies   1)-4). 
  In Fig. \ref{Fig2}.a we present the results
 versus  $\frac{M}{\sigma_l^2}$, where the cutoff frequency assumed by the graph CRB and by the estimator is set to $R=10$ and the number of sensors is $D=40$.
The
   random  sampling policy has been omitted from this figure since it has significantly high WMSE compared with the other methods.  
  It can be seen that  the proposed sampling allocation policy from Algorithm  \ref{Alg1} has  lower Dirichlet energy than that of the  E-design \cite{chen2015discrete} and A-design \cite{anis2016efficient} methods, 
  both in terms of the associated graph CRBs and the actual CML performance. 
  For a low noise variance shown in the figure, i.e. low  $\frac{1}{\sigma_l^2}$, the performance of the CML estimators deviated from the associated graph CRBs. This is due to the fact that in this figure we use real-data values of $\thetavec$, and, thus, $\thetavec$ is not a perfect bandlimited graph signal in practice. This mismatch prior assumption affects the performance, especially in the case of high SNRs. 
   In Fig.  \ref{Fig2}.b we present the results
 versus the number of selected sensors for a perfectly $R=10$ bandlimited graph signal and for $\sigma_l^2=0.5$. It can be seen that the proposed sampling allocation policy from Algorithm \ref{Alg1},  the A-design, and the E-design methods are all consistent, in the sense that the WMSE decreases with the number of samples, in contrast to the random sampling policy.
 The proposed method outperforms the other methods for any number of selected sensors.
These figures demonstrate the fact that the sampling set of $\Mmat$ affects the graph CRB differently for each choice of $D$ nodes. Therefore, by proper selection of sensors that are placed at the most informative locations (in the sense of the new gCRB), we can obtain the best performance for the same number of nodes, $R$. 

\begin{figure}[htb]
\subcaptionbox{\label{fig:mse01}}[\linewidth]
{\includegraphics[width=6.75cm]{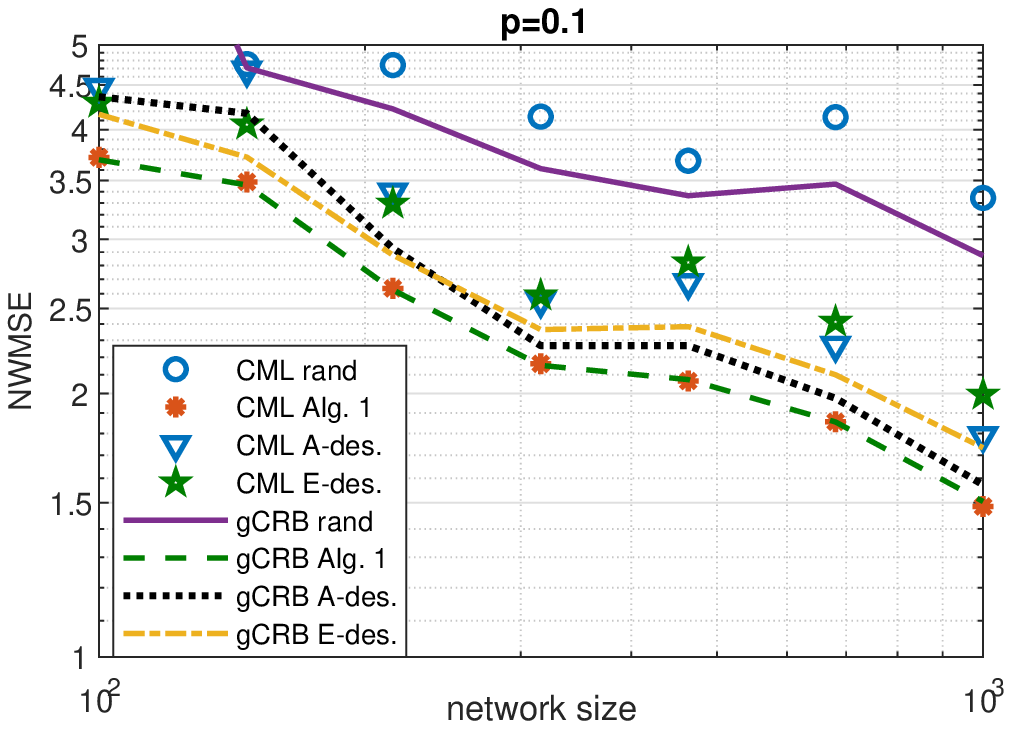}}
\subcaptionbox{\label{fig:mse02}}[\linewidth]
{\includegraphics[width=6.75cm]{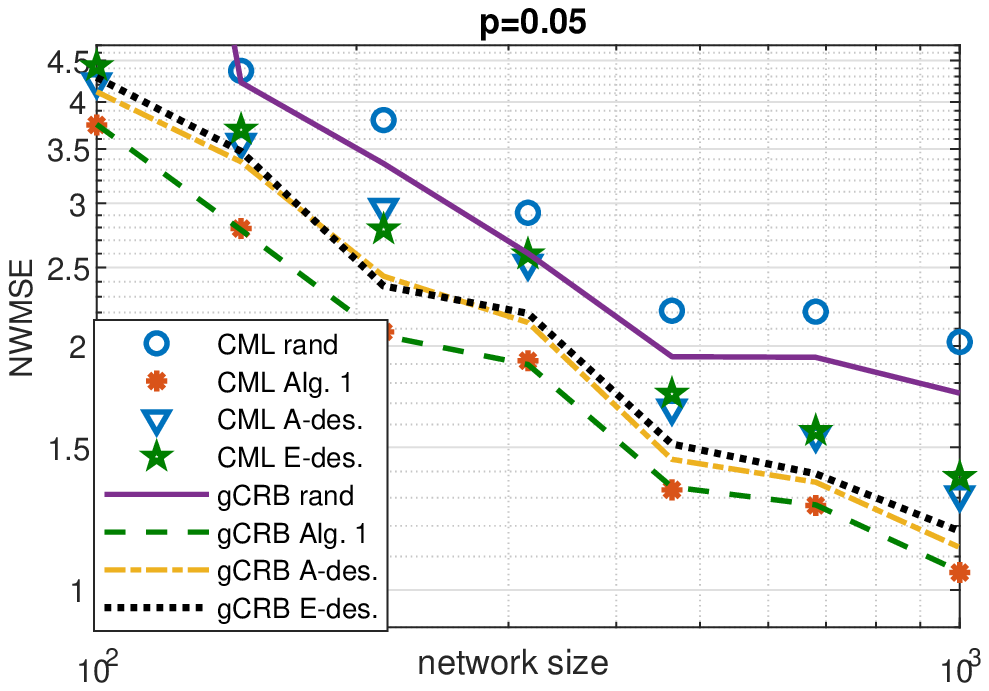}}
\subcaptionbox{\label{fig:runtime01}}[\linewidth]
{ \includegraphics[width=6.75cm]{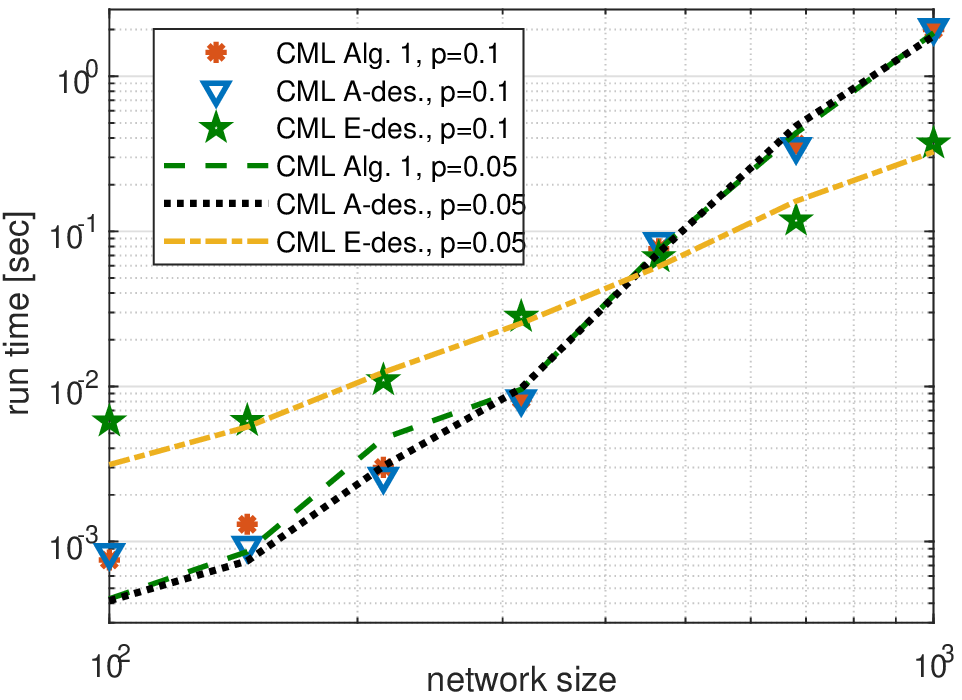}}
     \caption{Bandlimited graph signal recovery: Performance of the random sensor allocation policy, and Algorithm \ref{Alg1}, A-design, and E-design methods   for  Erd\H{o}s-R$\acute{\text{e}}$nyi graphs with edge connecting probabilities of (a) $0.1$, (b) $0.05$,
     o and (c) both $0.1$ and  $0.05$, 
     with $R=15$-bandlimited signals and $D=0.2 M$ sensors versus the number of nodes in terms of
     graph CRB and the root WMSE (a,b) and  runtime (c).}
     \label{Fig4}
      \vspace{-0.25cm}
\end{figure}

In Figs. \ref{Fig4}.a and \ref{Fig4}.b the  root WMSE
 of the CML estimator and the root of the graph CRB are presented for Erd\H{o}s-R$\acute{\text{e}}$nyi graphs
 versus the number of nodes in the system, $M$,
 with the sampling policies   1)-4).
 We consider the case of  $R=15$ graph frequencies 
 and $D=0.2 \times M$ sensors, and 
edge connecting probabilities $0.1$ and $0.05$.
 It can be seen that the proposed method outperforms the existing methods for any number of selected sensors and for the two edge connecting probabilities.
 The WMSE decreases as the network size decreases, as expected, since for all networks we have the same number of parameters to be estimated ($R$ parameters in the graph frequency domain),
 with an increasing number of sampled nodes, $D$. By comparing these figures, it can be seen that the estimation performance is better for less connected networks, i.e. the WMSE of all methods is lower for edge connecting probability of $p=0.05$.

In order to demonstrate the empirical complexity  of  Algorithm \ref{Alg1} for different network sizes and different graph topologies, 
 the average computation time was evaluated by running Algorithm \ref{Alg1} using Matlab on an Intel Core(TM) i7-7600U CPU computer, 2.80 GHz. 
Fig. \ref{fig:runtime01} shows the runtime of the different sampling policy methods versus $M$.  The runtime of the proposed Algorithm  \ref{Alg1} is identical to the runtime of the A-design method.
The E-design method has a higher runtime than the others for small networks, but is more scalable to large networks. However, its estimation performance 
 is worse
 than those of Algorithm  \ref{Alg1} and E-design methods. 
It can be seen that for all methods, as more  measurements are available,  the computation time increases since the search is over more values. However, the
edge connecting probability does not affect the runtime of the proposed approach.

	\section{Conclusion}
	\label{conclusion}
We consider the problem of graph signal recovery as a non-Bayesian parameter estimation  under the Laplacian-based WMSE performance evaluation measure. We develop the graph CRB, which is a lower bound on the WMSE of any graph-unbiased estimator in the Lehmann sense. We present new sensor allocation policies that aim to reduce the graph CRB under a constrained amount of sensing nodes. The relation between the problem of finding the optimal sensor locations of relative measurements in the graph-CRB sense and the problem of finding the maximum weight spanning tree of a graph  is demonstrated. The proposed graph CRB is evaluated and compared with the Laplacian-based WMSE of the CML estimator, for signal recovery over random graphs and for PSSE in electrical networks.
Significant performance 
gains are observed from these simulations for 
 using the optimal sensor locations. Thus, the proposed graph CRB can be used as a system design tool for  sensing networks.
 Future work includes extensions to  complex-valued systems and nonlinear examples,
 as well as developing low-complexity graph sampling methods based on the graph CRB that  avoid SVD computation that are similar to the approaches  in 
 \cite{wang2019low,bai2020fast}. In addition,
 methods  that include the cost of communication and computation should be developed, in order to assess the performance of distributive algorithms for graph signal recovery.

	\appendices
	\renewcommand{\thesectiondis}[2]{\Alph{section}:}
\section{Derivation of \eqref{Cmn3}}
\label{appA}
By using the facts that  $(\Vmat \Lambdamat^{\frac{1}{2}})^T=\Lambdamat^{\frac{1}{2}} \Vmat^T$ and that
$\Lambdamat^{\frac{1}{2}}$ is a diagonal matrix,
the $(m,n)$th element of the matrix cost function $\Cmat(\hat{\thetavec},\thetavec)$, defined in \eqref{Cmat_def}, satisfies
\beqna
\label{Cmn}
\Cmat_{m,n}(\hat{\thetavec},\thetavec)=\sum_{k=1}^M \sum_{l=1}^M
 [\Lambdamat^{\frac{1}{2}} \Vmat^T]_{m,k} [\Lambdamat^{\frac{1}{2}} \Vmat^T]_{n,l}\epsilon_k\epsilon_l \hspace{1cm}
\nonumber\\
= \Lambdamat^{\frac{1}{2}}_{m,m}\Lambdamat^{\frac{1}{2}}_{n,n}
\sum_{k=1}^M \sum_{l=1}^M 
\Vmat_{k,m} \Vmat_{l,n}\epsilon_k\epsilon_l \hspace{1.35cm}
\nonumber\\
=-\frac{1}{2} \Lambdamat^{\frac{1}{2}}_{m,m}  \Lambdamat^{\frac{1}{2}}_{n,n}\sum_{k=1}^M\sum_{l=1}^M
\Vmat_{k,m} \Vmat_{l,n}(\epsilon_k-\epsilon_l)^2
\nonumber\\
+\Lambdamat^{\frac{1}{2}}_{m,m}  \Lambdamat^{\frac{1}{2}}_{n,n}\sum_{k=1}^M 
\Vmat_{k,m}  \epsilon_k^2 \times \sum_{l=1}^M\Vmat_{l,n},\hspace{0.9cm}
\eeqna
where $\epsilon_m$, 
$m=1,\ldots,M$, are defined in \eqref{error_eps}.
Since $(\lambda_1,\vvec_1)=(0,\frac{1}{\sqrt{M}}\onevec)$ is  the 
smallest eigenvalue-eigenvector pair of the  positive
semidefinite Laplacian matrix, 
then, $\onevec$ is orthogonal to the other  $M-1$ eigenvectors of $\Lmat$ that are given by the last $M-1$ columns of 
$\Vmat$. Thus, $\sum\nolimits_{l=1}^M\Vmat_{l,n}=0$, $\forall n=2,\ldots,M$. Since in addition $ \Lambdamat^{\frac{1}{2}}_{1,1}=0$, we can conclude that
 \be
\label{eig2}
 \Lambdamat^{\frac{1}{2}}_{n,n}\sum\nolimits_{l=1}^M\Vmat_{l,n}=0,~\forall n=1,\ldots,M.
\ee 
By substituting \eqref{eig2} in 
\eqref{Cmn}, we obtain
\beqna
\label{Cmn2}
\Cmat_{m,n}(\hat{\thetavec},\thetavec)\hspace{5cm}
\nonumber\\=-\frac{1}{2} \Lambdamat^{\frac{1}{2}}_{m,m}  \Lambdamat^{\frac{1}{2}}_{n,n}\sum_{k=1}^M\sum_{l=1}^M
\Vmat_{k,m} \Vmat_{l,n}(\epsilon_k-\epsilon_l)^2.
\eeqna
By substituting
\beqna
\label{eee}
(\epsilon_k-\epsilon_l)^2=
(\epsilon_k-\epsilon_m+\epsilon_m-\epsilon_l+\epsilon_n-\epsilon_n)^2\hspace{1cm}
\nonumber\\
=(\epsilon_k-\epsilon_m)^2+(\epsilon_l-\epsilon_n)^2+(\epsilon_m-\epsilon_n)^2\hspace{0.7cm}
\nonumber\\
-2\left[(\epsilon_k-\epsilon_m)(\epsilon_l-\epsilon_n)
\right. \hspace{3.2cm}
\nonumber\\
\left.
+(\epsilon_k-\epsilon_m)(\epsilon_m-\epsilon_n)
+(\epsilon_l-\epsilon_n)(\epsilon_m-\epsilon_n)\right]
\eeqna
into \eqref{Cmn2} and reordering the elements, we obtain
\beqna
\label{Cmn2_think}
\Cmat_{m,n}(\hat{\thetavec},\thetavec)
= \Lambdamat^{\frac{1}{2}}_{m,m}  \Lambdamat^{\frac{1}{2}}_{n,n}\sum_{k=1}^M \Vmat_{k,m} \sum_{l=1}^M
\Vmat_{l,n}
\hspace{1.5cm}
\nonumber\\
\times\left\{-\frac{1}{2}
\left[(\epsilon_k-\epsilon_m)^2+(\epsilon_l-\epsilon_n)^2+(\epsilon_m-\epsilon_n)^2\right]
\right.\hspace{0.35cm}
\nonumber\\
+
\left[(\epsilon_k-\epsilon_m)(\epsilon_l-\epsilon_n)
\right. \hspace{4.2cm}
\nonumber\\
\left.
\left.
+(\epsilon_k-\epsilon_m)(\epsilon_m-\epsilon_n)
+(\epsilon_l-\epsilon_n)(\epsilon_m-\epsilon_n)\right]\right\}\hspace{0.8cm}
\nonumber\\
= \Lambdamat^{\frac{1}{2}}_{m,m}\sum_{k=1}^M\Vmat_{k,m} 
\left[-\frac{1}{2}(\epsilon_k-\epsilon_m)^2
\right.\hspace{2.45cm}
\nonumber\\
\left.
-\frac{1}{2}(\epsilon_m-\epsilon_n)^2+(\epsilon_k-\epsilon_m)(\epsilon_m-\epsilon_n)\right] \hspace{1.7cm}
\nonumber\\
\times
\left(\Lambdamat^{\frac{1}{2}}_{n,n}
\sum_{l=1}^M
\Vmat_{l,n}\right)\hspace{4.4cm}
\nonumber\\
+  \Lambdamat^{\frac{1}{2}}_{n,n}\sum_{l=1}^M \Vmat_{l,n} 
\left[-\frac{1}{2}(\epsilon_l-\epsilon_n)^2
+(\epsilon_l-\epsilon_n)(\epsilon_m-\epsilon_n)\right]\nonumber\\
\times \left(
\Lambdamat^{\frac{1}{2}}_{m,m}\sum_{k=1}^M\Vmat_{k,m}\right)\hspace{4.15cm}
\nonumber\\
+\Lambdamat^{\frac{1}{2}}_{m,m}  \Lambdamat^{\frac{1}{2}}_{n,n}\sum_{k=1}^M\sum_{l=1}^M
\Vmat_{k,m} \Vmat_{l,n} 
(\epsilon_k-\epsilon_m)(\epsilon_l-\epsilon_n).\hspace{0.24cm}
\eeqna
By substituting  \eqref{eig2}
in \eqref{Cmn2_think},
we obtain the term in \eqref{Cmn3}.

\section{Proof of Proposition \ref{propLehm}}
	\label{unbiasedApp}
	In this appendix, we prove that the graph-unbiasedness is obtained from the Lehmann unbiasedness definition in Definition \ref{unbiased_definition}
	with the cost function
	from \eqref{Cmat_def} and under the constrained set in \eqref{linear}.
	By substituting  (\ref{Cmat_def}) in (\ref{Lehmann_vector}),
 we obtain that the  Lehmann unbiasedness in this case requires that
\beqna 
\label{kMin}
\Lambdamat^{\frac{1}{2}} \Vmat^T{\rm{E}}_{\thetavecsmall}\left[(\hat{\thetavec}-\etavec)(\hat{\thetavec}-\etavec)^T\right]\Vmat
		\Lambdamat^{\frac{1}{2}}
		\hspace{2cm}
\nonumber\\
\succeq \Lambdamat^{\frac{1}{2}} \Vmat^T{\rm{E}}_{\thetavecsmall}\left[(\hat{\thetavec}-\thetavec)(\hat{\thetavec}-\thetavec)^T\right]\Vmat\Lambdamat^{\frac{1}{2}},
\eeqna
$\forall \thetavec,\etavec \in \Omega_\thetavecsmall$.
Similar to  the derivation on p. 14 of \cite{Lehmann},
on adding and subtracting
 ${\rm{E}}_{\thetavecsmall}[\hat{\thetavec}] $ inside the four round brackets in 
 \eqref{kMin},  this condition is
reduced to 
\beqna 
\label{kMin2}
\Lambdamat^{\frac{1}{2}} \Vmat^T
{\rm{E}}_{\thetavecsmall}\left[({\rm{E}}_{\thetavecsmall}[\hat{\thetavec}]-\etavec)
 ({\rm{E}}_{\thetavecsmall}[\hat{\thetavec}]-\etavec)^T\right]\Vmat\Lambdamat^{\frac{1}{2}}
\hspace{1.5cm}
\nonumber\\
\succeq \Lambdamat^{\frac{1}{2}} \Vmat^T{\rm{E}}_{\thetavecsmall}\left[({\rm{E}}_{\thetavecsmall}[\hat{\thetavec}]-\thetavec) ({\rm{E}}_{\thetavecsmall}[\hat{\thetavec}]-\thetavec)^T\right]\Vmat\Lambdamat^{\frac{1}{2}},
\eeqna
 $\forall \thetavec, \etavec \in\Omega_\thetavecsmall$. By using the definition of the constrained set in \eqref{set_def},
and since the range of $\Umat$  is the null-space of $\Gmat$,
	it can be seen that for a given  $\thetavec\in\Omega_\thetavecsmall$,
	any $\etavec\in\Omega_\thetavecsmall$ can be written as (see, e.g. Section 4.2.4 in \cite{Boyd_2004})
	\be
	\label{w_vec}
	\etavec=\thetavec+\Umat\wvec,
	\ee
	for some vector $\wvec\in {\mathbb{R}}^{M-K}$.
	By substituting \eqref{w_vec} in \eqref{kMin2}, we obtain
	\beqna 
\label{kMin3}
\Lambdamat^{\frac{1}{2}} \Vmat^T
{\rm{E}}_{\thetavecsmall}\left[({\rm{E}}_{\thetavecsmall}[\hat{\thetavec}]-\thetavec-\Umat\wvec)
 ({\rm{E}}_{\thetavecsmall}[\hat{\thetavec}]-\thetavec-\Umat\wvec)^T\right]\Vmat\Lambdamat^{\frac{1}{2}}
\nonumber\\
\succeq \Lambdamat^{\frac{1}{2}} \Vmat^T{\rm{E}}_{\thetavecsmall}\left[({\rm{E}}_{\thetavecsmall}[\hat{\thetavec}]-\thetavec) ({\rm{E}}_{\thetavecsmall}[\hat{\thetavec}]-\thetavec)^T\right]\Vmat\Lambdamat^{\frac{1}{2}},
\eeqna
	$\forall \thetavec\in\Omega_\thetavecsmall, \wvec\in {\mathbb{R}}^{M-K}$.
	The condition in \eqref{kMin3} can be rewritten as 
		\beqna 
\label{kMin4}
\Lambdamat^{\frac{1}{2}} \Vmat^T\Umat\wvec\wvec^T\Umat^T\Vmat\Lambdamat^{\frac{1}{2}}\succeq 
\hspace{4.25cm}
\nonumber\\
\Lambdamat^{\frac{1}{2}} \Vmat^T
\left(
({\rm{E}}_{\thetavecsmall}[\hat{\thetavec}]-\thetavec
)\wvec^T\Umat^T
+\Umat\wvec({\rm{E}}_{\thetavecsmall}[\hat{\thetavec}]-\thetavec
)^T
\right)\Vmat\Lambdamat^{\frac{1}{2}}.
\eeqna
A necessary condition for  \eqref{kMin4} to be satisfied is that
\beqna
\label{kMin5}
\wvec^T\Umat^T\Lmat\Umat\wvec
\geq 2  \wvec^T\Umat^T\Lmat
({\rm{E}}_{\thetavecsmall}[\hat{\thetavec}]-\thetavec
),
\eeqna
where we applied the trace operator on both sides on \eqref{kMin4}  and used the trace operator properties.
Since the l.h.s. of \eqref{kMin5} is nonnegative, 
(\ref{unbiased_cond}) is a sufficient condition for \eqref{kMin5} to hold. 
In addition,
since the condition in \eqref{kMin5} should be satisfied for any $\wvec\in {\mathbb{R}}^{M-K}$, it should be satisfied
	  in particular for both $\wvec =  \epsilon \evec_k\in {\mathbb{R}}^{M-K}$ and $\wvec =  -\epsilon \evec_k\in {\mathbb{R}}^{M-K}$, $k=1,\ldots,K$, with small positive $\epsilon$.  By substituting these values in  \eqref{kMin5},
	 we require
	 \beqna
\label{kMin6}
\epsilon^2 \evec_k^T\Umat^T\Lmat\Umat\evec_k
\geq  \pm \epsilon 2 \evec_k^T\Umat^T\Lmat
({\rm{E}}_{\thetavecsmall}[\hat{\thetavec}]-\thetavec
),
\eeqna
for any $k=1,\ldots,K$, which, by taking the limit $\epsilon\rightarrow 0^+$, implies that 
	 $[\Umat^T\Lmat
({\rm{E}}_{\thetavecsmall}[\hat{\thetavec}]-\thetavec
)]_k=0$ for any $k=1,\ldots,K$, and, thus,
		(\ref{unbiased_cond}) is also 
	a necessary condition for \eqref{kMin5}
 to be satisfied. Thus, we can conclude that (\ref{unbiased_cond})  is a sufficient condition for
 the  graph unbiasedness  of estimators in the Lehmann sense in \eqref{kMin4}.

	\section{Proof of Theorem \ref{T3}}
	\label{App_T3}
	The following proof for the development of the graph CRB is along the path of the development 
of the  CCRB on the MSE in a conventional estimation problem in \cite{Stoica_Ng}.
Let $\Wmat\in{\mathbb{R}}^{M\times M}$  be an arbitrary
matrix.  Then, the Cauchy-Schwarz inequality implies that
	\beqna
	\label{Wmat_one}
	\Lambdamat^{\frac{1}{2}} \Vmat^T{\rm{E}}_\thetavecsmall\left[\left(\hat{\thetavec}-\thetavec-\Wmat\Umat\Umat^T\nabla_{\thetavecsmall}^T\log f(\xvec;\thetavec)\right)
	\right.\hspace{1.25cm}
	\nonumber\\
	\times
	\left.
	\left(\hat{\thetavec}-\thetavec-\Wmat\Umat\Umat^T\nabla_{\thetavecsmall}^T\log f(\xvec;\thetavec)\right)^T\right]\Vmat \Lambdamat^{\frac{1}{2}}
	\succeq\zerovec.
\eeqna
Under  Condition \ref{cond1},
the matrix inequality in \eqref{Wmat_one} implies that
\beqna
\label{one1}
	\Lambdamat^{\frac{1}{2}} \Vmat^T{\rm{E}}_\thetavecsmall[(\hat{\thetavec}-\thetavec)(\hat{\thetavec}-\thetavec)^T]
	\Vmat \Lambdamat^{\frac{1}{2}}\hspace{3.2cm}
	\nonumber\\
	\succeq
	\Lambdamat^{\frac{1}{2}} \Vmat^T{\rm{E}}_\thetavecsmall[(\hat{\thetavec}-\thetavec) \nabla_{\thetavecsmall}\log f(\xvec;\thetavec)]\Umat\Umat^T\Wmat^T\Vmat \Lambdamat^{\frac{1}{2}}\hspace{0.2cm}
		\nonumber\\
	+\Lambdamat^{\frac{1}{2}} \Vmat^T\Wmat\Umat\Umat^T{\rm{E}}_\thetavecsmall[\nabla_{\thetavecsmall}^T\log f(\xvec;\thetavec)(\hat{\thetavec}-\thetavec)^T]\Vmat \Lambdamat^{\frac{1}{2}}
		\nonumber\\
	-\Lambdamat^{\frac{1}{2}} \Vmat^T\Wmat\Umat\Umat^T\Jmat(\thetavec)\Umat\Umat^T\Wmat^T\Vmat \Lambdamat^{\frac{1}{2}}.\hspace{1.625cm}
	\eeqna
By using regularity condition \ref{cond2} with $g(\xvec,\thetavec)=
(\hat{\thetavec}-\thetavec)  f(\xvec;\thetavec)
$ and the product rule, it can be verified that (see, e.g. Appendix 3B in \cite{Kayestimation})
\beqna
\label{reg0}
{\rm{E}}_\thetavecsmall\left[ (\hat{\thetavec}-\thetavec) \nabla_{\thetavecsmall}\log f(\xvec;\thetavec)\right]
=\Imat_M+\nabla_{\thetavecsmall} {\rm{E}}_\thetavecsmall[ \hat{\thetavec}-\thetavec].
\eeqna
Thus, by multiplying \eqref{reg0} by $\Lmat$
and $\Umat$ from left and right, respectively,
and substituting the local graph-unbiasedness from \eqref{sec_localU_final} in Definition \ref{LocCunbias_prop}, we obtain
\beqna
\label{reg022}
\Lmat {\rm{E}}_\thetavecsmall\left[ (\hat{\thetavec}-\thetavec) \nabla_{\thetavecsmall}\log f(\xvec;\thetavec)\right]\Umat
=\Lmat\Umat.
\eeqna
Or, equivalently, by using $\Lmat=\Vmat\Lambdamat \Vmat^T$ and the fact that $\Lambdamat$ is a nonnegative diagonal matrix, 
\beqna
\label{reg02}
\Lambdamat^{\frac{1}{2}} \Vmat^T {\rm{E}}_\thetavecsmall\left[ (\hat{\thetavec}-\thetavec) \nabla_{\thetavecsmall}\log f(\xvec;\thetavec)\right]\Umat
=\Lambdamat^{\frac{1}{2}} \Vmat^T\Umat.
\eeqna
By  substituting  \eqref{MSSE} and \eqref{reg02} in \eqref{one1}, we obtain
	\beqna
	\label{two2}
{\rm{E}}_{\thetavecsmall}[\Cmat(\hat{\thetavec},\thetavec)]	\succeq
	\Lambdamat^{\frac{1}{2}} \Vmat^T\Umat\Umat^T\Wmat^T\Vmat \Lambdamat^{\frac{1}{2}}
	\hspace{2.25cm}
	\nonumber\\
	+\Lambdamat^{\frac{1}{2}} \Vmat^T\Wmat\Umat\Umat^T\Vmat \Lambdamat^{\frac{1}{2}}
	\hspace{2.25cm}
		\nonumber\\
	-\Lambdamat^{\frac{1}{2}} \Vmat^T\Wmat\Umat\Umat^T\Jmat(\thetavec)\Umat\Umat^T\Wmat^T\Vmat \Lambdamat^{\frac{1}{2}}.
\eeqna
Now, let $\Wmat$ be such that
\beqna
\label{www}
\Wmat\Umat=\Umat\left(\Umat^T\Jmat(\thetavec)\Umat\right)^\dagger.
\eeqna
By substituting \eqref{www} in \eqref{two2} and using $\Amat^\dagger\Amat\Amat^\dagger=\Amat^\dagger$ for any matrix $\Amat$, we obtain
the bound in \eqref{CR_bound1}-\eqref{CR_bound2}.

According to the conditions for equality in the Cauchy-Schwarz inequality, for any given matrix $\Wmat$,
 the equality  holds in \eqref{Wmat_one} {\em{iff}}
\beqna
\label{four}
\Lambdamat^{\frac{1}{2}} \Vmat^T (\hat{\thetavec}-\thetavec)=\zeta(\thetavec) \Lambdamat^{\frac{1}{2}} \Vmat^T \Wmat\Umat\Umat^T\nabla_{\thetavecsmall}^T\log f(\xvec;\thetavec),
\eeqna
where $\zeta(\thetavec)$ is a scalar function of $\thetavec$.
By substituting \eqref{www} in \eqref{four}, we obtain the following condition
 for  achievability:
\beqna
\label{five}
\Lambdamat^{\frac{1}{2}} \Vmat^T (\hat{\thetavec}-\thetavec)=\hspace{4.5cm}
\nonumber\\\zeta(\thetavec)\Lambdamat^{\frac{1}{2}} \Vmat^T \Umat\left(\Umat^T\Jmat(\thetavec)\Umat\right)^\dagger\Umat^T\nabla_{\thetavecsmall}^T\log f(\xvec;\thetavec).
\eeqna
Computing the expected quadratic term ($\avec\avec^T$) of each side of \eqref{five}, 
substituting \eqref{MSSE} and \eqref{JJJdef},  we obtain
\beqna
\label{Su_equality_cond_preview_W11}
{\rm{E}}_{\thetavecsmall}[\Cmat(\hat{\thetavec},\thetavec)]
=\zeta^2(\thetavec) \Bmat(\thetavec),
\eeqna
where $\Bmat(\thetavec)$ is defined in \eqref{CR_bound2}.
 Thus, for obtaining equality in \eqref{CR_bound1}, we require $\zeta(\thetavec)=\pm 1$. 
It can be verified that in order for $\hat{\thetavec}$ 
from \eqref{Su_equality_cond_preview_W11} to satisfy \eqref{sec_localU_final} from Definition \ref{LocCunbias_prop}, we must take 
$
\zeta(\thetavec)= 1$.
By substituting  $
\zeta(\thetavec)= 1$ in \eqref{five}, we obtain \eqref{equality_cond}.

\section{Derivation of \eqref{J_trans_inv4}}
\label{pinv_app}
	Since  $\bar{\Emat}\bar{\Emat}^T$ is a Laplacian matrix, it satisfies 
	$\onevec^T\left(\bar{\Emat}\bar{\Emat}^T-\frac{1}{M}\onevec\onevec^T\right)^{-1}=-\onevec^T$ and $\left(\bar{\Emat}\bar{\Emat}^T-\frac{1}{M}\onevec\onevec^T\right)^{-1}\onevec=-\onevec$. By using these results,
	the null-space property of $\Lmat$, and $\bar{\Lmat}=\bar{\Lmat}-\frac{1}{M}\onevec\onevec^T+\frac{1}{M}\onevec\onevec^T$, we obtain that  \eqref{J_trans} can be rewritten as
		\beqna
	\label{J_trans_inv}
	\Jmat(\thetavec)=	\frac{1}{\sigma^2}
\hspace{6cm}\nonumber\\
			\times
		\left(\bar{\Lmat}-\frac{1}{M}\onevec\onevec^T\right)\left(\bar{\Emat}\bar{\Emat}^T-\frac{1}{M}\onevec\onevec^T\right)^{-1}\left(\bar{\Lmat}-\frac{1}{M}\onevec\onevec^T\right).
	\eeqna
By using Lemma 3 from \cite{boley2011commute}, we obtain
	\beqna
	\label{J_trans_inv2}
	\Jmat^\dagger(\thetavec)=	
	\sigma^2(\Imat-\frac{\psivec\psivec^T}{\psivec^T \psivec})
		\left(\bar{\Lmat}-\frac{1}{M}\onevec\onevec^T\right)^{-1}\hspace{1.5cm}\nonumber\\
		\times\left(\bar{\Emat}\bar{\Emat}^T-\frac{1}{M}\onevec\onevec^T\right)\left(\bar{\Lmat}-\frac{1}{M}\onevec\onevec^T\right)^{-1}\left(\Imat-\frac{\yvec\yvec^T}{\yvec^T \yvec}\right),
	\eeqna
	where
$	\psivec
		=-\frac{1}{\sqrt{M}}\onevec$
	$\yvec^T
	=\frac{1}{\sqrt{M}}\onevec^T$,
	and we used the facts that
	$\onevec^T\left(\bar{\Emat}\bar{\Emat}^T-\frac{1}{M}\onevec\onevec^T\right)^{-1}=-\onevec^T$ and $\left(\bar{\Emat}\bar{\Emat}^T-\frac{1}{M}\onevec\onevec^T\right)^{-1}\onevec=-\onevec$.
	By substituting $\psivec$ and $\yvec$
	in \eqref{J_trans_inv2}, one obtains
		\beqna
	\label{J_trans_inv3}
	\Jmat^\dagger(\thetavec)=	
	\sigma^2(\Imat-\frac{1}{M}\onevec\onevec^T)
		\left(\bar{\Lmat}-\frac{1}{M}\onevec\onevec^T\right)^{-1}\hspace{1cm}\nonumber\\
		\times\left(\bar{\Emat}\bar{\Emat}^T-\frac{1}{M}\onevec\onevec^T\right)\left(\bar{\Lmat}-\frac{1}{M}\onevec\onevec^T\right)^{-1}(\Imat-\frac{1}{M}\onevec\onevec^T).
\eeqna
It is well known  that the pseudo-inverse of a Laplacian matrix with $M$ nodes is given by
\cite{GUTMAN_Xiao_2004}
\be
\label{Linv}
\bar{\Lmat}^\dagger=\left(\bar{\Lmat}-\frac{1}{M}\onevec\onevec^T\right)^{-1}+\frac{1}{M}\onevec\onevec^T.
\ee
By substituting
\eqref{Linv} in \eqref{J_trans_inv3} and using the null-space property of $\bar{\Emat}\bar{\Emat}^T$ and $\bar{\Lmat}^\dagger$, we obtain \eqref{J_trans_inv4}.


\small
\bibliographystyle{IEEEtran}

\end{document}